\documentclass[
twocolumn,
amsfonts,amsmath,amssymb,
aps,
prd,
floatfix,
showpacs,
longbibliography,
nofootinbib
]{revtex4-2}
\usepackage{hyperref}
\usepackage[normalem]{ulem}
\usepackage{graphicx}
\usepackage{bm}
\usepackage[dvipsnames]{xcolor}
\usepackage{color}

\definecolor{ReflexBlue}{rgb}{ .0902,.0902,.5882}
\hypersetup{
    colorlinks=true,
    linkcolor=ReflexBlue,
    filecolor=magenta,      
    urlcolor=ReflexBlue,
    citecolor=ReflexBlue,
}
\let\Phi\varPhi

\newcommand{\Rs}{R_{\mathrm{S}}}
\newcommand{\beq}{\begin{equation}}
\newcommand{\eeq}{\end{equation}}

\definecolor{CPcolor}{rgb}{0.2, 0.13, 0.9}

\definecolor{PBcolor}{rgb}{0.9, 0.5, 0.0}

\newcommand{\EE}{{\mathcal{E}}}
\newcommand{\PP}{{\mathcal{P}}}
\newcommand{\ZZ}{{\mathcal{Z}}}
\newcommand{\TT}{{\mathcal{T}}}
\newcommand{\pt}{p_\perp}
\newcommand{\con}{\mathcal{C}}
\begin{document}

\title{Slowly rotating covariant anisotropic objects}
\author{Philip Beltracchi}
\email{phipbel@aol.com}
\affiliation{}
\author{Camilo Posada}
\email{camoposada82@gmail.com}
\affiliation{Programa de Matem\'atica, Fundaci\'on Universitaria Konrad Lorenz, 110231 Bogot\'a, Colombia}

\begin{abstract}
The equilibrium configurations of slowly rotating anisotropic self-gravitating fluids are computed using the extended Hartle structure equations, including anisotropic effects, derived in our previous paper. We focus on the so-called $\con$-star, whose anisotropic pressure follows a fully covariant equation of state (EOS), while a standard polytrope describes the radial pressure. We determine surface and integral properties, such as the moment of inertia, mass change, mass quadrupole moment, and ellipticity. Notably, for certain values of the compactness parameter, highly anisotropic $\con$-stars exhibit a prolate shape rather than the typical oblate form {\textemdash}an intriguing behavior also observed in other anisotropic systems like Bowers-Liang spheres and stars governed by a quasilocal EOS. Although the $\con$-stars considered in this study are limited by stability criteria and cannot sustain compactness beyond $M/R\approx0.38$, we found indications that certain rotational perturbations exhibit similarities to those observed in other ultracompact systems approaching the black hole limit.
\end{abstract}

\maketitle

\section{Introduction}

In a recent paper~\cite{Beltracchi:2024dfb}, we extended Hartle's second-order framework in angular frequency for slowly rotating relativistic stars~\cite{Hartle:1967he, Hartle:1968si} to incorporate anisotropic configurations within general relativity. Various approaches have been developed to study rotating systems in general relativity, each with its own advantages and limitations. For instance, the Newman-Janis algorithm~\cite{Newman:1965tw,Newman:1965my,Wiltshire:2009zza}, which was instrumental in the discovery of the Kerr-Newman metric, has been generalized in several ways~\cite{Gurses:1975vu,Drake:1998gf,Azreg-Ainou:2014pra} to generate rotating versions of spherically symmetric spacetimes. While these extensions of the Newman-Janis algorithm provide exact solutions to the Einstein field equations, they generally do not preserve the equation of state (EOS) between the static and rotating configurations~\cite{Beltracchi:2021ris, Beltracchi:2021tcx}.

Additionally, several perturbative approaches have been developed to model slowly rotating systems. The Lense-Thirring metric describes frame-dragging effects in weak gravitational fields~\cite{Lense:1918zz}, while the Hartle-Thorne~\cite{Hartle:1967he, Hartle:1968si} method extends this framework to study the second-order in rotation structural changes in rotating objects obeying a perfect fluid EOS. Our work further generalizes this formalism to include systems with more general EOSs~\cite{Beltracchi:2024dfb}. As an application of this formalism, we examined the structure and integral properties of slowly rotating Bowers-Liang spheres~\cite{Beltracchi:2024dfb}, which are characterized by uniform density and anisotropic pressure. While the Bowers-Liang sphere provides a simple, closed-form analytic solution, its physical relevance is limited by the assumption of constant density and the \emph{ad hoc} nature of its anisotropic EOS.

Although neutron stars (NSs) are usually modeled with isotropic pressure, it has been proposed that various physical phenomena can lead to local anisotropies; for instance, strong magnetic fields~\cite{Yazadjiev:2011ks,Frieben:2012dz,Folomeev:2015aua}, pion condensation~\cite{Sawyer:1972cq}, and scalars fields minimally coupled to gravity~\cite{Liebling:2012fv}, among others~\cite{Herrera:1997plx}. The effect of anisotropy on the NSs properties has been studied in a number of papers~\cite{Bowers:1974tgi,Bayin:1982vw, 
Silva:2014fca,Yagi:2015hda,Biswas:2019gkw,Das:2022ell}. In this context, several methods to introduce anisotropies have been proposed in the literature. Cosenza \emph{et al.}~\cite{Cosenza:1981myi} proposed an anisotropic term proportional to the gravitational force. A quasilocal EOS for the anisotropic pressures, where the anisotropy is proportional to the compactness and the radial pressure, was proposed by Ref.~\cite{Horvat:2010xf}. Additional methods include the complexity factor approach~\cite{Herrera:2018bww}, the minimal geometric deformation~\cite{Ovalle:2019lbs}, and the covariant method proposed by Ref.~\cite{Raposo:2018rjn}. More recently, Refs.~\cite{Alho:2021sli, Alho:2023mfc} proposed relativistic elastic stars as possible models of ultracompact objects and black hole (BH) mimickers.

Among the various methods discussed above for introducing anisotropies, the covariant approach proposed by Ref.~\cite{Raposo:2018rjn}, or $\con$-star, offers a viable alternative for obtaining a rotating solution while preserving the EOS. In this model, the anisotropy is proportional to the directional derivative of the pressure, a certain arbitrary function of the mass density, and a parameter $\con$ that quantifies the `strength' of the anisotropy. For specific parameter choices of the solution, \cite{Raposo:2018rjn} found that $\con$-stars are stable, have vanishing tidal Love numbers, and produce gravitational wave echoes when perturbed.

In this paper, we employ the extended Hartle structure equations derived in Ref.~\cite{Beltracchi:2024dfb} (see also Ref.~\cite{Becerra:2024xff}), to study the structure and integral properties of slowly rotating $\con$-stars, up to second order in angular velocity. We consider a $\con$-star model in which the anisotropic EOS follows the covariant approach outlined above, while the EOS relating radial pressure to mass density corresponds to a standard polytrope. Thus, our analysis extends the work of Ref.~\cite{Miller:1977} on slowly rotating isotropic polytropes. 

Recently, the authors of Ref.~\cite{Becerra:2025snm} reported a short study of $\con$-stars in slow rotation. However, they restricted their analysis to investigating the impact of the anisotropy on the mass, radius, binding energy, and moment of inertia. Our analysis complements and extends the one in Ref.~\cite{Becerra:2025snm} in several ways: first of all, we compute all the monopole and quadrupole perturbation functions for $\con$-stars, up to order $\Omega^2$ in rotation. We also determine the surface and integral properties, such as the moment of inertia, the change of mass, the ellipticity, and the mass quadrupole moment. 

The structure of the paper is as follows. in Sec.~\ref{sect:hartle}, we review the extended formalism for slowly rotating anisotropic relativistic stars. In Sec.~\ref{sect:cstars}, we summarize the covariant model for anisotropic relativistic fluids, or $\con$-stars. The rotational properties, as well as structure and integral properties of $\con$-stars, are presented in Sec.~\ref{sect:results}. Finally, in Sec.~\ref{sect:disc}, we discuss our final conclusions and open questions for future work. In Appendix~\ref{appendixA}, we provide more details on the behavior of the $\con$-star near the origin. In Appendix~\ref{appendix B}, we examine the correspondence between the covariant relativistic conservation equation and classical force densities to further analyze the prolate deformation. Throughout the paper, we use geometrized units, where $c=G=1$, unless stated otherwise.

\section{Review of the slowly rotating anisotropic Hartle formalism}
\label{sect:hartle}
The structure equations for slowly rotating anisotropic configurations were already derived in \cite{Beltracchi:2024dfb}, therefore, we refer the reader to that reference for further details. Here, we summarize the formalism and main results.
\subsection{Rotational metric perturbations and energy-momentum tensor are specified}
The standard Hartle perturbative line element for slowly rotating systems reads~\cite{Hartle:1967he,Hartle:1968si}
\begin{align}
&ds^2 =  - e^{2\nu_0(r)} \Big[ 1+2 h_0(r) + 2 h_2(r)\, P_2(\cos\theta)  \Big]  dt^2 \nonumber\\ & 
 +e^{2\lambda_0(r)}\left\{1+\frac{2 e^{2\lambda_0(r)}}{r}\Big[ m_0(r) + m_2(r)\, P_2(\cos\theta) \Big] \right\}  dr^2 \nonumber\\& 
+ r^2 \Big[ 1+2 k_2(r)\, P_2(\cos\theta)  \Big] \Big[ d\theta^2 + \sin^2\!\theta\,\big(d\phi - \omega(r) dt \big)^2 \Big].
\label{kingmet}
\end{align}
Here, $\nu_0(r)$ and $\lambda_0(r)$ are the metric functions associated to the nonrotating solution, $h_l(r)$, $k_l(r)$, and $m_l(r)$, are the monopole ($l=0$) and quadrupole ($l=2$) perturbation functions of order $\Omega^2$, and $P_2(\cos\theta)$ indicates the Legendre polynomial of order $l=2$. The function $\omega(r)$, associated with the inertial \emph{frame dragging}, is of the first order in $\Omega$. Finally, the condition $k_0(r)=0$ corresponds to Hartle's gauge~\cite{Hartle:1967he}.

The energy-momentum tensor (EMT) for a rotating anisotropic configuration takes the general form
\beq
T^\mu_{~\nu}=(\mathcal{E}+\mathcal{T})u^\mu u_\nu+\mathcal{T} \delta^\mu_{~\nu}-(\mathcal{T}-\mathcal{P}) v^\mu v_\nu, \label{anisotropicEMTcov}
\eeq
where $\EE$ denotes the energy density, $\PP$ is the radial pressure, $\TT$ is the transverse pressure in the comoving frame of the rotating fluid, $u$ is a normalized four-velocity, and $v$ is a normalized vector along the spacelike eigenvector with a distinct eigenvalue. The eigenvalue of Eq.~\eqref{anisotropicEMTcov}, namely, $\mathcal{E}$, $\mathcal{P}$, and $\mathcal{T}$ have the degeneracy behavior of the Segre type [(11)1,1]~\cite{Beltracchi:2024dfb}. Following~Hartle~\cite{Hartle:1967he}, we expand the set of eigenvalues in the following fashion: 
\beq\label{eigenE}
\EE = \rho(r) + \EE_0(r) + \EE_2(r) P_2(\cos\theta),
\eeq
\beq\label{eigenP}
\PP = p_r(r) + \PP_0(r) + \PP_2(r) P_2(\cos\theta),
\eeq
\beq\label{eigenT}
\TT = p_\perp(r)+\TT_0(r)+\TT_2(r)P_2(\cos\theta),
\eeq
where we have zeroth-order terms and second-order monopole and quadrupole terms in rotation.

The components of the four-velocity $u^{\mu}$ take the standard Hartle form~\cite{Hartle:1967he,Hartle:1968si}
\begin{align}
u^t &= \left(-g_{tt}-2\Omega g_{t\phi}-\Omega^2 g_{\phi \phi}\right)^{-1/2}, \nonumber\\
u^\phi &= \Omega u^t,\quad u^r = u^\theta=0,
\label{4vhartle}
\end{align}
and \cite{Beltracchi:2024dfb}
\beq
v^\mu=\left(0,\frac{1}{\sqrt{g_{rr}}},Y\Omega^2,0\right),
\label{kdef}
\eeq
with 
\beq
(p_r-\pt)Y=\Upsilon(r)\sin\theta\cos\theta, 
\label{ups_first}
\eeq
in $(t,r,\theta,\phi)$ coordinates.
\subsection{A nonrotating configuration is computed}
To the zeroth order in $\Omega$, Eq.~\eqref{kingmet} reduces to a spherically symmetric line element in the standard Schwarzschild form
\beq\label{metric_stat}
ds^2 = -e^{2\nu_0(r)}dt^2 + e^{2\lambda_0(r)}dr^2 + r^2 (d\theta^2 + \sin^2 d\phi^2),
\eeq
where $\nu_0$ and $\lambda_0$ are functions of the radial coordinate $r$. In the static case, the expansions of the eigenvalues of the EMT, Eqs.~\eqref{eigenE}-\eqref{eigenT}, reduce to their values at rest. Thus, for a static and spherically symmetric configuration, the general form~\eqref{anisotropicEMTcov} reduces to
\beq\label{emt}
T^{\mu}_{\,\,\nu}=\mathrm{diag}\left(-\rho, ~p_{r},~p_{\perp},~p_{\perp}\right),
\eeq
\noindent where $\rho$ is the energy density, $p_r$ is the radial pressure, and $p_{\perp}=p_{\theta}=p_{\phi}$ is the transverse pressure. In the isotropic case, $p_\perp=p_r$. 

For a given value of the central density, the nonrotating solution is determined by integrating the anisotropic Tolman-Oppenheimer-Volkoff (ATOV) equation of hydrostatic equilibrium for the radial pressure $p_r$~\cite{Beltracchi:2024dfb}
\beq\label{atov}
\frac{dp_r}{dr}=-\left(\rho + p_{r}\right)\frac{(m+4 \pi r^3 p_r)}{r(r-2m)} + \frac{2}{r}(p_{\perp}-p_r),
\eeq
and the equation for the total mass $m(r)$ enclosed inside a spherical surface of radius $r$, as follows
\beq\label{mass}
\frac{dm}{dr}=4\pi\rho(r)r^2.
\eeq
The mass $m(r)$ is connected to the $g_{rr}$ metric component, via the standard relation
\beq\label{grrmass}
e^{-2\lambda_0}=1-\frac{2m(r)}{r}.
\eeq
The total mass of the nonrotating background configuration $M=m(R)$, where $R$ is the stellar radius. In the isotropic case, Eq.~\eqref{atov} reduces to the standard TOV equation. 

The system \eqref{atov}-\eqref{grrmass} consists of three equations with five unknowns, requiring two additional EOSs to relate $\rho$, $p_r$, and $\pt$. Once the EOSs are specified, the integration is performed from the center of the star with the initial conditions $m(0)=0$, $\rho(0)=\rho_c$ (where $\rho_c$ is the central density), and $p(\rho_c)=p_c$, where the central pressure $p_c$ is given by the corresponding EOS. The radius of the configuration is determined by imposing the boundary condition $p_r(R)=0$.

The remaining metric function $\nu_0(r)$ can be obtained by integrating outward from the origin the equation
\beq
\frac{d\nu_0}{dr}=\frac{(m+4 \pi r^3 p_r)}{r(r-2m)},
\label{dnudr}
\eeq
with the boundary condition $e^{2\nu_0(R)}=1-2M/R$, from the matching at the boundary with the exterior Schwarzschild solution.

\subsubsection{Exterior spacetime and matching conditions}
In the exterior vacuum spacetime $\rho=p=0$, thus the spacetime geometry is described by Schwarzschild's exterior solution
\beq
e^{2\nu_0(r)}=e^{-2\lambda_0(r)}=1-\frac{2M}{r},\quad r>R.
\eeq
Unless there is a junction layer at the surface, the interior and exterior geometries are matched at the boundary $\Sigma=R$, such that
\beq
[\nu_0]=0,\quad [\nu_0']=0,\quad [\lambda_0]=0,
\eeq
where the prime denotes derivative with respect to the $r$ coordinate, and $[f]$ indicates the difference between the value of $f$ in the vacuum exterior and its value in the interior, evaluated at $\Sigma$, i.e., $[f]=f^{+}\vert_{\Sigma}-f^{-}\vert_{\Sigma}$.

\subsection{Frame dragging and moment of inertia}
The $t\phi$, or $\phi t$, components of the Einstein equations are first order in $\Omega$. Defining the functions 
\beq
\varpi\equiv \Omega-\omega,    
\eeq
and 
\beq
j\equiv e^{-(\nu_0+\lambda_0)}=e^{-\nu_0}\sqrt{1-\frac{2m}{r}},    
\eeq
the frame-dragging equation can be written as \cite{Beltracchi:2024dfb} (see also Ref.~\cite{Pattersons:2021lci}) 
\begin{align}
\frac{1}{r^4}\frac{d}{dr}\left(r^4 j \frac{d\varpi}{dr}\right)+\frac{4j'\varpi}{r}\frac{\rho+\pt}{\rho+p_r}=0. \label{newomega}
\end{align}
In the isotropic case, where $p_r=\pt$, Eq.~\eqref{newomega} reduces to the original Hartle relation~\cite{Hartle:1967he}. In the exterior region $r>R$, $\rho=p=0$ and $j(r)=1$, thus Eq.~\eqref{newomega} can be solved explicitly to give
\begin{equation}\label{omegaout}
\varpi(r)^{+}=\Omega - \frac{2J}{r^3},
\end{equation}
\noindent where $J$ is an integration constant associated with the angular momentum of the star \cite{Hartle:1967he}. To find the interior solution, Eq.~\eqref{newomega} must be integrated outward from the origin with the boundary conditions
\begin{subequations}
\beq
\varpi(0)=\varpi_{\mathrm{c}}=\mathrm{const}.,
\eeq
\beq
\left(\frac{d\varpi}{dr}\right)_{r=0}=0.
\eeq
\label{bc_omega}
\end{subequations}
We require regularity of the solution at the surface, such that $[\varpi]=[\varpi']=0$~\footnote{Hartle~\cite{Hartle:1967he} chose $[\varpi]=0$ by construction. However, it is possible to match an interior and exterior spacetime such that the values $\varpi$ takes, as one approaches the matching surface from the inside and outside, are different, if one has suitable definitions to relate the interior and exterior coordinates at the matching surface. However, we choose Hartle's gauge for the interior region where $\varpi$ is continuous, in consequence $\Omega$ corresponds to the angular velocity of the fluid as measured by a distant observer~\cite{Reina:2014fga,Reina:2015jia}.}. To determine the angular momentum $J$ of the configuration, and the angular velocity $\Omega$, Eq.~\eqref{newomega} must be numerically integrated, thus one can obtain the surface value $\varpi(R)$. Then $J$ and $\Omega$ can be obtained from the following relations: 
\begin{equation}\label{wsurf}
J=\frac{1}{6}R^4\left(\frac{d\varpi}{dr}\right)_{r=R},\quad \Omega = \varpi(R) + \frac{2J}{R^3}.
\end{equation}
Subsequently, the relativistic moment of inertia is obtained from the relation $I = J/\Omega$.
\subsection{The monopole perturbations: the \texorpdfstring{$l=0$}{l=0} sector}
There are three differential equations for the monopole sector. The one for the perturbation function $m_0$ comes from the $tt$ component of Einstein's equation~\cite{Beltracchi:2024dfb}
\beq
\frac{dm_0}{dr} = 4\pi r^2 \EE_0 +\frac{j^2 r^4}{12}\left(\frac{d\varpi}{dr}\right)^2 - \frac{r^3\varpi^2}{3}\frac{dj^2}{dr}\frac{\rho+\pt}{\rho+p_r}. 
\label{newm0}
\eeq
The $rr$ perturbation equation results in a differential equation for $h_0$ which takes the form~\cite{Beltracchi:2024dfb}
\beq
\frac{dh_0}{dr} = 4\pi r e^{2\lambda}\PP_0+\frac{(1+8\pi r^2 p_r)}{(r-2m)^2}m_0-\frac{e^{2\lambda}r^3 j^2 \varpi'^2}{12}.
\label{newh0}
\eeq
Finally, using $(8\pi T^\theta_{~\theta}-G^\theta_{~\theta})+(8\pi T^\phi_{~\phi}-G^\phi_{~\phi})=0$, or the monopole perturbation to the $r$ component of the EMT conservation $\nabla_\mu T^\mu_{~~\nu}=0$, we obtain~\cite{Beltracchi:2024dfb} 
\begin{multline}
\frac{d\PP_0}{dr}=-(\EE_0+\PP_0)\nu_0'+\frac{2(\TT_0-\PP_0)}{r}+\frac{\rho+\pt}{3}\frac{d}{dr}\left(\frac{r^3\varpi^2j^2}{r-2m}\right)\\
+\frac{(\rho+p_r)}{r-2m}\left[\frac{r^4j^2\varpi'^2}{12} - 4\pi r^2 \PP_0 - \frac{\left(1+8\pi r^2 p_r\right)}{(r-2m)}m_0\right].
\label{newp0}
\end{multline}
These equations are integrated outward, from the origin, where $m_0(0)=\PP_0(0)=0$.
\subsubsection{Matching and setup for the monopole sector}
It is convenient to define the auxiliary function
\beq
H_0\equiv h_0-h_c,
\label{hc_eqn}
\eeq
where $h_c$ is a constant to be determined. Note that $H_0$ satisfies the differential equation~\eqref{newh0}, thus it must be integrated with the initial condition $H_0(0)=0$. To find the constant $h_c$, we impose continuity of the solution, i.e, the values of $H_0(r\rightarrow R^-)$ and the exterior $h_0(r\rightarrow R^+)$, since the background solution $g_{tt}$ is nonzero.

Outside the configuration where $\rho=p=0$, Eqs.~\eqref{newm0} and \eqref{newh0} can be integrated explicitly, giving
\beq
m_0^{+}=\delta M-\frac{J^2}{r^3},
\label{m0out}
\eeq
\beq
h_0^{+}=-\frac{m_0}{r-2M},
\label{h0out}
\eeq
where $\delta M$ is a constant of integration associated with the change of mass induced by the rotation.

Hartle~\cite{Hartle:1967he} determined the constant $\delta M$ by assuming the continuity of $m_0$ and $h_0$, at the surface, i.e., $[h_0]=[m_0]=0$. However, Refs.~\cite{Reina:2014fga, Reina:2015jia} found that for systems where the density at the surface is nonvanishing, there appears a discontinuity in $m_0$ which generates an additional term in Eq.~\eqref{m0out}. In the anisotropic case, we showed that the modified change of mass takes the general form~\cite{Beltracchi:2024dfb}
\beq
\delta M_{\mathrm{mod}}=m_0(R)+\frac{J^2}{R^3}+4\pi\rho(R)R^2 \xi_{0},
\label{dmMODfin}
\eeq
where $\xi_0$ is the spherical deformation parameter
\begin{align}
    \xi_0=-\frac{\PP_0}{p_r'(r\rightarrow R^-)}.
    \label{xi0def}
\end{align}
 However, this extra term is unnecessary in our case since the EOSs we consider guarantee that the density vanishes at the surface.

\subsection{The quadrupole perturbations: the \texorpdfstring{$l=2$}{l=2} sector}

\label{Sec:quad}
The quadrupole sector is dependent on the three-metric perturbation functions $h_2,k_2,m_2$, three EMT eigenvalue perturbation functions $\EE_2,\PP_2,\TT_2$, and an EMT eigenvector perturbation function $\Upsilon$ for a total of seven functions (in contrast to five in the isotropic case) \cite{Beltracchi:2024dfb}. One equation comes from the Einstein equations $r\theta$ component
\begin{multline}
    8\pi r^3 \Upsilon(r)=3 e^{-\lambda} \Bigg[r 
   \frac{d}{dr}\left(h_2+k_2\right)+h_2 \left(r
 \frac{d\nu_0}{dr}  -1\right)\\-\frac{m_2}{r-2m} \left(r \frac{d\nu_0}{dr}+1\right)\Bigg].
 \label{upsdef}
\end{multline} 
Additionally, the $\theta$ component of the EMT conservation equation gives
\begin{multline}
3\TT_2+(\rho+p_\perp)\left(3 h_2+ e^{-2\nu_0}\varpi^2 r^2\right) =\\ 
(p_r-p_\perp)\left(\frac{3e^{2\lambda}}{r}\right)m_2 + e^{-\lambda}\left[r\Upsilon(4+r\nu_0') + r^2 \Upsilon'\right].
\label{constheta}
\end{multline}
Einstein's equations require $(8\pi T^\theta_{~\theta}-G^\theta_{~\theta})-(8\pi T^\phi_{~\phi}-G^\phi_{~\phi})=0$, the expansion of which which gives an algebraic equation which could eliminate $m_2$ or $h_2$, as follows:
\begin{equation}
    m_2=(r-2m)\left[\frac{j^2 r^4 \varpi'^2}{6}-\frac{r^3\varpi^2}{3}\left(\frac{dj^2}{dr}\right)\frac{\rho+\pt}{\rho+pr}-h_2\right].\label{m2item}
\end{equation}
Using Eq.~\eqref{m2item} to replace $m_2$, the $rr$ component of the Einstein field equations becomes
\begin{widetext}
\begin{multline}
\frac{j^2 r^2}{3}\left[16\pi(\rho+\pt)(1+2r\nu_0')\varpi^2+\left(\frac{3}{2}+8\pi r^2p_r\right)\left(\frac{d\varpi}{dr}\right)^2\right]=
-8\pi \PP_2+\frac{2(r-2m)}{r^2}\left[\frac{dh_2}{dr}+(1+r\nu_0')\frac{dk_2}{dr}\right] \\
+\frac{4}{r^2}\left[h_2(4\pi r^2 p_r-1)-k_2\right].
\label{quadA}
\end{multline}
\end{widetext}

Finally, the quadrupole perturbation to the $r$ component of $\nabla_\mu T^\mu_{~~\nu}=0$ gives
\begin{widetext}
\begin{multline}
\frac{d\PP_2}{dr}=\frac{2(\TT_2-\PP_2)}{r}-(\EE_2+\PP_2)\nu_0'+\frac{2}{3}r\varpi e^{-2\nu_0}(\rho+\pt)\left[2\varpi(1+2r\nu_0')-r\left(\frac{d\varpi}{dr}\right)\right]+\frac{j^2 r(1+r\nu_0')}{8\pi}\left(\frac{d\varpi}{dr}\right)^2 \\
-\left(\frac{3}{4\pi r^2}\right)\frac{d}{dr}\left(h_2+k_2\right)-(\rho+p_r)\frac{dh_2}{dr}+2(\pt-p_r)\frac{dk_2}{dr}-\left(\frac{3\nu_0'}{2\pi r^2}\right)h_2.
\label{quadC}
\end{multline}
\end{widetext}
The system of Eqs.~\eqref{upsdef}--\eqref{quadC}, plus two EOSs defines the seven equations that determine the quadrupole sector.
\subsubsection{Matching and setup for the quadrupole sector}
In the vacuum exterior spacetime, the functions $h_2$ and $k_2$ are found to be~\cite{Hartle:1967he,Hartle:1968si}
\beq\label{h2out}
h_{2}=\frac{J^2}{Mr^3}\left(1 + \frac{M}{r}\right) + KQ_{2}^{\;2}(\zeta),
\eeq
\beq\label{v2out}
v_{2}=h_{2}+k_{2}=-\frac{J^2}{r^4} + K\frac{2M}{\left[r(r-2M)\right]^{1/2}}Q_{2}^{\;1}(\zeta),
\eeq
\noindent where $K$ is an integration constant, and $Q_{n}^{\;m}$ are the associated Legendre functions of the second kind with argument $\zeta\equiv (r/M) -1$, which read (following the convention in Ref.~\cite{Hartle:1967he})
\begin{subequations}
\begin{align}
&Q_{2}^{\;1}(\zeta)=\sqrt{\zeta^2-1}~\left[\frac{3\zeta^2 -2}{\zeta^2-1}-\frac{3}{2}\zeta\log\left(\frac{\zeta+1}{\zeta-1}\right)\right],\\
&Q_{2}^{\;2}(\zeta)=\frac{3}{2}\left(\zeta^2 -1\right)\log\left(\frac{\zeta+1}{\zeta-1}\right)-\frac{3\zeta^3-5\zeta}{\zeta^2-1}.
\end{align}
\end{subequations}

To match with the exterior solution, one solves the system in the interior, Eqs.~\eqref{upsdef}--\eqref{quadC}, for both the particular case with nonzero $\varpi$ and the homogeneous case with $\varpi=0$, then choosing the true functions~\footnote{Although the authors of Ref.~\cite{Becerra:2025snm} seem to have derived the same quadrupole equations, they do not describe the procedure of using the homogeneous solution to match the exterior solution at the outer boundary, as it was done in Ref.~\cite{Beltracchi:2024dfb}. This is crucial for solving the quadrupole sector.}
\beq
\text{true} = \text{particular} + A*\text{homogeneous},
\label{split}
\eeq
where $A$ is the same constant for all the functions. The integration constants $A$ and $K$ are determined by matching $h_2$ and $k_2$ to the exterior solutions given by Eqs.~\eqref{h2out} and \eqref{v2out}. Once the constant $K$ has been computed, the mass quadrupole moment of the star, as measured at infinity, can be obtained from the relation~\cite{Hartle:1968si}
\beq
Q=\frac{J^2}{M}+\frac{8}{5}K M^3.
\eeq
Following Hartle~\cite{Hartle:1967he}, the EMT eigenvalue perturbations $\EE_2,\PP_2,\TT_2$ vanish at the origin. The function $\Upsilon$, which describes the eigenvector perturbation, also goes to zero at the origin since it is proportional to a factor of $p_r-p_\perp$, which is zero there.

For well-behaved systems, the initial behavior of the relevant zeroth and first-order terms in $\Omega$ is as follows~\cite{Beltracchi:2024dfb}: 
\begin{align}
m &= \frac{4\pi \rho_c}{3}r^3,\\
\nu_0 &= \nu_c +\frac{2\pi}{3}(\rho_c + 3p_c)r^2,\\
p_r &= p_\perp=p_c,\\
\rho &= \rho_c,\\
\varpi &= \varpi_c,
\end{align}
where the subscript ‘$c$' denotes the central value of the quantity. We found series solutions near the origin for the particular Eqs.~\eqref{upsdef}, \eqref{constheta}, \eqref{quadA}, and \eqref{quadC}, as follows \cite{Beltracchi:2024dfb}:
\beq
k_{2}^{\text{(P)}} = k_2^a r^2+...+k_2^b r^4,
\label{k2p_expa}
\eeq
\beq 
h_{2}^{\text{(P)}} = -k_2^a r^2+...+h_2^b r^4,
\label{h2p_expa}
\eeq
The coefficients to the leading quadratic terms for the $h_2$ and $k_2$ metric functions must be equal and opposite. The lowest order terms in the $\Upsilon,\PP_2,\TT_2$ functions are related to the quadratic and quartic, but not cubic,\footnote{The $\con$-star exhibits cubic terms in the expansions of the quadrupole functions [see Eq.~\eqref{quadcube}] a feature absent in the Bowers-Liang sphere~\cite{Beltracchi:2024dfb}.} coefficients of the $h_2$ and $k_2$ functions:
\begin{multline}    
\Upsilon^{\text{(P)}} = \Bigg[\frac{3(h_2^b+k_2^b)}{2\pi}-e^{-2\nu_c}\varpi_c^2(p_c+\rho_c)-\\
k_2^a(3p_c+\rho_c)\Bigg]r,
\label{yp_expa}
\end{multline}
\begin{multline}
\PP_{2}^{\text{(P)}} = \frac{1}{6}\Bigg[\frac{3(h_2^b+k_2^b)}{\pi}+4k_2^a \rho_c\\
-4e^{-2\nu_c}\varpi_c^2(p_c+\rho_c)\Bigg]r^2,
\label{p2p_expa}
\end{multline}
\begin{multline}
\TT_{2}^{\text{(P)}} = \Bigg[\frac{5(h_2^b+k_2^b)}{2\pi}-2e^{-2\nu_c}\varpi_c^2(p_c+\rho_c)\\
-\frac{2}{3}k_2^a(6p_c+\rho_c)\Bigg]r^2,    
\label{T2p_expa}
\end{multline}
where $h_{2}^{b}$, $k_{2}^{a}$, and $k_{2}^{b}$ are arbitrary constants.
\subsubsection{Deformation of the star}
The rotational deformation of the star can be understood in the following way: a given constant $p_r$ surface, lying at some radius $r$ in the static configuration, upon rotation, is displaced to a radius
\beq
r+\xi_0(r) + \xi_2(r)P_2(\cos\theta),
\eeq
where $\xi_0$ is the monopole deformation given by Eq.~\eqref{xi0def}, and the perturbation function $\xi_2$ is defined as
\beq
\xi_2=-\frac{\PP_2}{p_r'(r\rightarrow R^-)}.
\label{xi2def}
\eeq    
The ellipticity of the isobaric surfaces is given by~\cite{Beltracchi:2024dfb}
\beq
\epsilon(r)=-\frac{3}{2r}\left[\xi_2(r) + rk_2(r)\right].
\label{elipticitydef}
\eeq
The ellipticity of the stellar surface is computed by setting $r=R$ in Eq.~\eqref{elipticitydef}.

\section{$\mathcal{C}$-stars}
\label{sect:cstars}
We consider the covariant approach for anisotropic configurations proposed in Ref.~\cite{Raposo:2018rjn}, which introduces a covariant framework to satisfy the condition at the center, $p_r=\pt$. Although there is a certain degree of arbitrariness to fulfill this condition, Ref.~\cite{Raposo:2018rjn} chose the following ansatz
\beq\label{raposoeos}
p_\perp=p_r-\con f(\rho) v^\mu \nabla_\mu p_r,
\eeq
where $f(\rho)$ is a function of the density, $v^\mu$ is the corresponding EMT eigenvector to $p_r$, and $\con$ is a constant that quantifies the strength of the anisotropy. Note that, by construction, $p_r=\pt$ at the center for the static spherically symmetric case, since $\partial_r p_r (0)=0$.

Considering the ansatz \eqref{raposoeos}, the ATOV equation~\eqref{atov} takes the form
\beq
\frac{dp_r}{dr}=-\frac{(\rho+p_r)}{r(r-2m)}\frac{(m+4\pi p_r r^3)}{1+\frac{2}{r}\con f(\rho)\sqrt{1-\frac{2m}{r}}}.
\eeq
We recover the standard TOV equation in the isotropic case $\con=0$.
In order to close the system, we require two EOSs for $p_r$ and $f(\rho)$. Following Ref.~\cite{Raposo:2018rjn}, we choose 
\beq
f(\rho)=\rho\label{EOSfdef},
\eeq
which ensures the continuity of $\pt$ and its derivative at the surface; for the radial pressure we choose a standard polytropic EOS 
\beq
p_r= \kappa\rho_0^\gamma,\label{polytropdef}
\eeq
where $\rho_0=\rho-p_r/(\gamma-1)$ denotes the rest-mass density, $\kappa$ and $\gamma=1+1/n$ are constants for any particular model ($n$ is the polytropic index). 
There are a few properties that led us to examine this EOS, or $\con$-star, for our rotating framework. First, it can be naturally expressed in terms of scalars, making it well suited for generalization to the rotating case. In contrast, other promising methods in the spherically symmetric case, such as the quasilocal EOS~\cite{Horvat:2010xf}, where the anisotropy is proportional to the compactness $m(r)/r$, is not formed from terms that we found an obvious way to isolate as scalars in the rotating case. Nevertheless, the quasilocal EOS has been extended to slow rotation~\cite{Becerra:2024xff} by writing the EOS, in the rotating case, in terms of the nonrotating compactness function. However, the authors in Ref.~\cite{Becerra:2024xff} did not elaborate on whether or not this is a valid approximation.

For this system, we can use the EOS to write the background Eqs.~\eqref{atov} and \eqref{mass} in terms of $p_r$ and $m$, as follows: 
\beq
\frac{dm}{dr}=4\pi r^2\left[\frac{p_r}{\gamma-1}+\left(\frac{p_r}{\kappa}\right)^{1/\gamma}\right]~ \label{cstardmpr},\\
\eeq
\begin{multline}
\frac{dp_r}{dr}=-\frac{\left(m+4\pi p_r r^3\right) \left[\left(p_r/\kappa\right)^{1/\gamma}+(1+n)p_r\right]}{r(r-2m)}\\
\times\left\{1+\frac{2\con}{r}\left[\left(\frac{p_r}{\kappa}\right)^{1/\gamma}+np_r\right]\sqrt{1-\frac{2m}{r}}\right\}^{-1},
\label{cstarTOVpr}
\end{multline}
or in terms of $\rho_0$ and $m$ as 
\beq\label{dmrho0}
\frac{dm}{dr} = 4\pi r^2\left(\rho_0+\frac{\kappa\rho_0^{\gamma}}{\gamma-1}\right),
\eeq
\begin{multline}
\frac{d\rho_0}{dr} = -\left(\frac{\rho_0^{-1/n}}{\gamma\kappa}\right)\\
\times\frac{\left[\rho_0+\kappa(1+n)\rho_0^\gamma\right]\left(m+4\pi r^3 \kappa \rho_0^\gamma \right)}{r(r-2m)\left[1+\frac{2\con}{r}\left(\rho_0+\frac{\kappa\rho_0^{\gamma}}{\gamma-1}\right)\sqrt{1-\frac{2m}{r}}\right]}.
\label{drho0dr}
\end{multline}  
In our approach, we consider the $(\rho_0,m)$ representation because it gives the most stable numerical results.~
\subsection{Equation of state and the perturbations}
We carry out the generalization to the rotating case with $\rho\rightarrow\mathcal{E}$, $p_r\rightarrow\mathcal{P}$, and $p_\perp\rightarrow\mathcal{T}$. Notice that since these are EMT eigenvalues, the covariant derivative term becomes the partial derivative $\partial_\mu P$. 

It is possible to rewrite the EOSs \eqref{raposoeos}, \eqref{EOSfdef}, and \eqref{polytropdef}, purely in terms of $\mathcal{P}$ and expand to get the expressions
\begin{align}
    \mathcal{E}&=\left(\frac{\mathcal{P}}{\kappa}\right)^{1/\gamma}+\frac{\mathcal{P}}{\gamma-1} ,\\
    \mathcal{T}&=\mathcal{P}-\con\left[\left(\frac{\mathcal{P}}{\kappa}\right)^{1/\gamma}+\frac{\mathcal{P}}{\gamma-1}\right]v^\mu \partial_\mu\mathcal{P},~
\end{align}
then, 

\beq
\mathcal{E}_x=\left(\frac{d\mathcal{E}}{d\mathcal{P}}\right)\mathcal{P}_x
=\left[\frac{(\frac{p_r}{\kappa})^{1/\gamma}}{\gamma p_r}+\frac{1}{\gamma-1}\right]\mathcal{P}_x,\label{expx}
\eeq
where the subscript $x$ takes the values $0,2$, corresponding to the monopole and quadrupole perturbation orders. Explicitly, using the definition of $v^\mu$ from Eq.~\eqref{kdef}, we can also compute the expressions for $\mathcal{T}_x$ as follows: 

\begin{align}
   \mathcal{T}_x&=\mathcal{P}_x -\frac{\con
   \left[(\gamma -1)
   \left(\frac{p_r}{\kappa}\right)^{\frac{1}{\gamma
   }}+\gamma p_r\right]p_r'}{\gamma(\gamma -1) 
   \sqrt{\frac{r}{r-2 m}} p_r}\mathcal{P}_x\nonumber\\ &
   +\frac{\con \left[(\gamma -1)
   \left(\frac{p_r}{\kappa}\right)^{\frac{1}{\gamma
   }}+p_r\right] 
   \left[m_x
   p_r'-(r-2 m)
   \mathcal{P}_x'\right]}{(\gamma-1)\sqrt{r (r-2
   m)}}.\label{txpx}
\end{align}
Alternatively, one can introduce a rotating generalization to the rest mass density, in the following form 
\beq
\mathcal{Z}=\rho_0+\mathcal{Z}_0+P_2(\cos\theta)\mathcal{Z}_2,
\eeq
where $\mathcal{Z}_0$ and $\mathcal{Z}_2$ are second order in $\Omega$~, and then write 
\beq\label{PPvsZZ}
\mathcal{P}_x=\kappa\gamma \rho_0^{\gamma-1} \mathcal{Z}_x,
\eeq
\beq\label{EEvsZZ}
\mathcal{E}_x=\left(1+\frac{\kappa\gamma \rho_0^{\gamma-1}}{\gamma-1}\right)\mathcal{Z}_x,
\eeq
\begin{align}
\mathcal{T}_x&= \gamma \kappa \rho_0^{\gamma-1} \mathcal{Z}_x \nonumber\\&
-\frac{\con \gamma \kappa \rho_0^{\gamma-1} \rho_0' \left[\gamma(\gamma -1)
    +(2\gamma-1)\kappa\rho_0^{\gamma
   -1}\right]}{(\gamma-1)\sqrt{\frac{r}{r-2m}}}\mathcal{Z}_x 
   \nonumber\\&
   +\frac{\con \gamma \kappa \rho_0^{\gamma }
   \left(\gamma-1+\kappa\rho_0^{\gamma-1}\right)
   \left[\rho'_{0} m_x - (r-2m)\mathcal{Z}'_x\right]}{(\gamma -1) \sqrt{r(r-2m)}}
\label{TTvsZZ}.
\end{align}

\section{Numerical Results}
\label{sect:results}
The extended Hartle structure equations, summarized in Sec.~\ref{sect:hartle}, were numerically integrated using methods similar to those employed in Ref.~\cite{Beltracchi:2024dfb} to study anisotropic homogeneous Bowers-Liang spheres in slow rotation. In contrast to the homogeneous model, the EOSs considered here do not permit an analytic solution describing the nonrotating configuration. Thus, for a given triplet ($n, \kappa, \con$), we determined the bounding radius $R$ and total mass $M(R)$ by numerically integrating the hydrostatic equilibrium equation \eqref{atov}, specifically using formulations \eqref{dmrho0} and \eqref{drho0dr}, varying the central rest-mass density $\rho_{0}(0)=\rho_{0_c}$ as the only free parameter of the solution. The stellar surface $R$ was found by the condition $p_r(R)=0$. Once the total mass and radius were obtained, we computed the configuration's compactness, $GM/(c^2 R)$.

The central value of the metric function $e^{\nu_0}$ was determined by numerically integrating Eq.~\eqref{dnudr} using, either the shooting method [or defining an auxiliary function analogously to the ($h_0$, $H_0$) construction in our consistency check codes] and setting the matching condition at the boundary with the exterior Schwarzschild metric. Once the nonrotating configuration was fully determined, i.e., $R$, $M(R)$, $\nu_0(r)$, and $\lambda_0(r)$ are known for the background solution, we solved for the monopole and quadrupole perturbation functions by simultaneously integrating the extended structure equations with their respective initial conditions. For instance, for the monopole sector we integrated simultaneously the coupled ordinary differential equations (ODEs) system, Eqs.~\eqref{newomega}, \eqref{newm0}, \eqref{newh0}, and \eqref{newp0}. Similarly, for the quadrupole sector, we integrated the system \eqref{upsdef}--\eqref{quadC} in the representation $(h_2,k_2,\ZZ_2,\Upsilon)$. The principal results of the numerical integrations are illustrated in Figs.~\ref{fig:mass-rad}--\ref{fig:q}. We introduce dimensionless variables, with units specified in the corresponding figure descriptions. Although the central rest-mass density $\rho_{0_c}$ is the only free parameter of the solution, we plotted the various perturbation functions against the compactness, to facilitate the comparison with some of the results reported by Ref.~\cite{Miller:1977} for slowly rotating isotropic polytropes.

As a consistency check, we also constructed additional codes to solve the background sector and import the background quantities as \texttt{Mathematica}'s~\cite{Mathematica} “interpolating functions" for solving the perturbations. For the monopole sector, we implemented these codes for the EOSs \eqref{expx} and \eqref{txpx} in terms of $\PP_0$, and $\ZZ_0$. For the quadrupole sector, we constructed multiple backup codes, one solving four ODEs in terms of the functions $(h_2, k_2, \ZZ_2, \Upsilon)$, another solving four ODEs in terms of the functions $(h_2, k_2, \PP_2, \Upsilon)$, and a third reducing the system to second-order equations involving only $h_2$ and $k_2$.

\subsection{Mass-radius diagram of $\con$-stars}

\begin{figure}[h]
\includegraphics[width=\columnwidth]{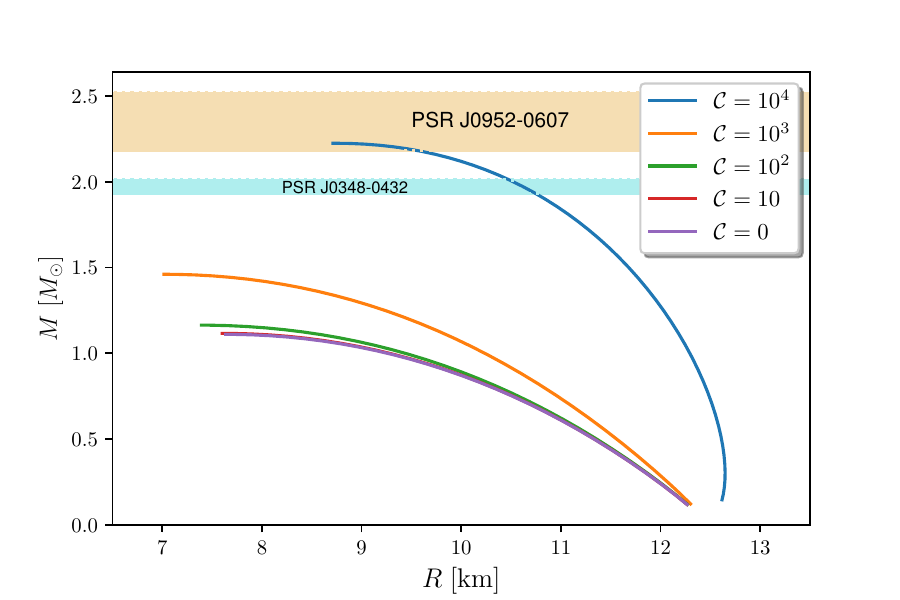}
\caption{Mass-radius diagram for $\con$-stars, for different values of the anisotropy parameter $\con$. We consider a polytrope with $\kappa=100$ and $n=1$. The maximum of the curves indicates the onset of instability. The color filled horizontal bands indicate the observational mass constraints from the pulsars PSR J0952-0607 \cite{Romani:2022jhd} and PSR J0348-0432 \cite{Demorest:2010bx}.}
\label{fig:mass-rad}
\end{figure}

A sequence of equilibrium $\con$-star configurations, with fixed values of the parameters $n$, $\kappa$, and $\con$, can be constructed by integrating the ATOV system \eqref{dmrho0} and \eqref{drho0dr}, varying the central rest-mass density $\rho_{0_c}$ as a free parameter. Following this procedure, we obtained the mass-radius diagram for $\con$-stars, shown in Fig.~\ref{fig:mass-rad}, for various values of the anisotropy parameter $\con$. Henceforth we focus on a configuration with $\kappa=100$, and $n=1$, which reproduces the reference configuration, in the isotropic case, presented in Ref.~\cite{Kokkotas:2000up}. The maximum of each curve marks the onset of instability, as determined by the standard $M(R)$ method~\cite{Shapiro:1983du,Horvat:2010xf}. In all subsequent computations, we restrict ourselves to configurations with $\rho_{0\,c}$ below the critical value for stability.

\begin{figure}[h]
\includegraphics[width=\columnwidth]{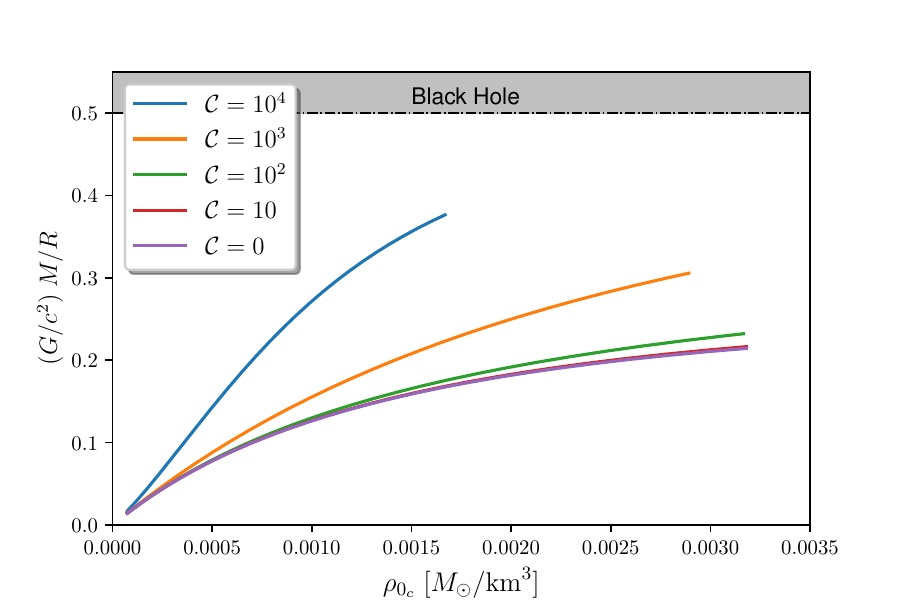}    
\caption{Compactness $GM/c^2 R$, as a function of the central density, for the same values of the anisotropic parameter $\con$ as in Fig.~\ref{fig:mass-rad}. Note that, for a given value of the central density, as the anisotropy increases, the compactness also increases.}
\label{fig:rho-comp}
\end{figure}

In Fig.~\ref{fig:rho-comp}, we show the compactness $GM/(c^2 R)$, as a function of $\rho_{0_c}$, for different values of the anisotropy parameter $\con$. The integration was halted at the critical value of $\rho_{0_c}$ for the onset of instability. Notably, the maximum compactness we found for radial stability in the isotropic case ($\con =0$), closely matches the value reported by Ref.~\cite{Miller:1977}, which was obtained through a static stability analysis of isentropic gas spheres. Note that for a given value of the central rest-mass density, an increase in the anisotropy increases the compactness. 

\subsection{Frame dragging and moment of inertia}

To determine the dragging of the inertial frames in $\con$-stars, we numerically integrated Eq.~\eqref{newomega} with the boundary conditions \eqref{bc_omega}, starting from the origin (or rather from some cutoff value $r_0=10^{-6}$), up the surface $R$. This was done for various configurations and different values of the anisotropy parameter $\con$.

\begin{figure*}
\centering
\includegraphics[width=.495\linewidth]{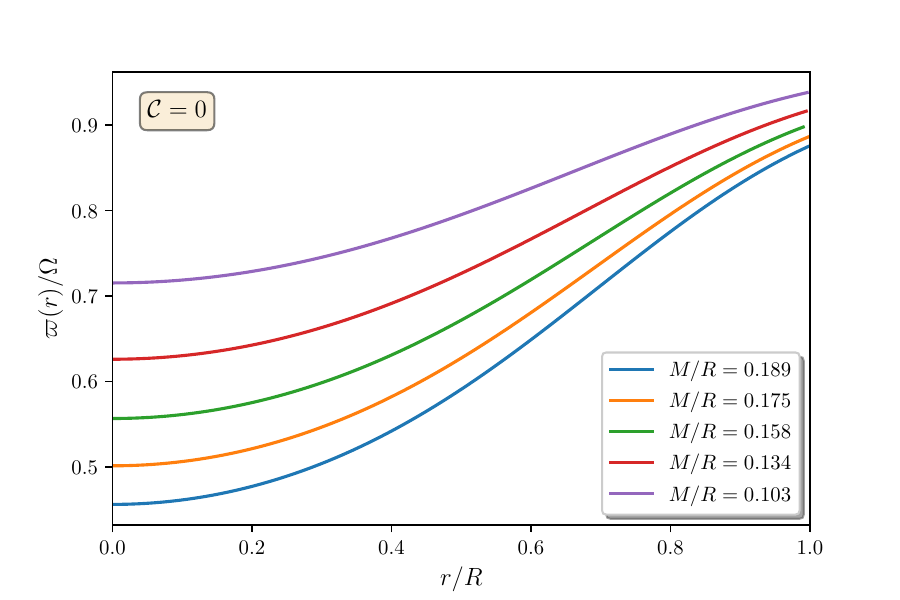}
\includegraphics[width=.495\linewidth]{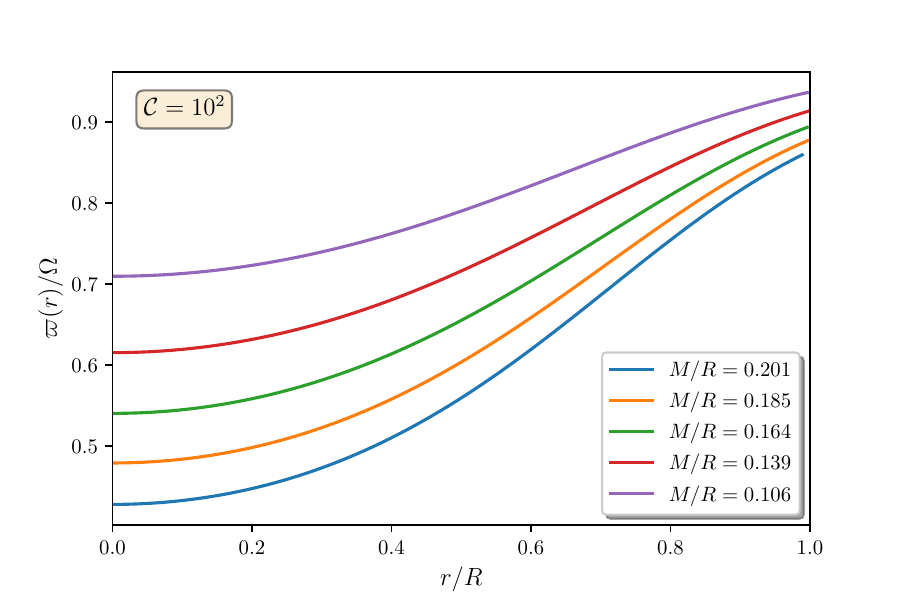}
\includegraphics[width=.495\linewidth]{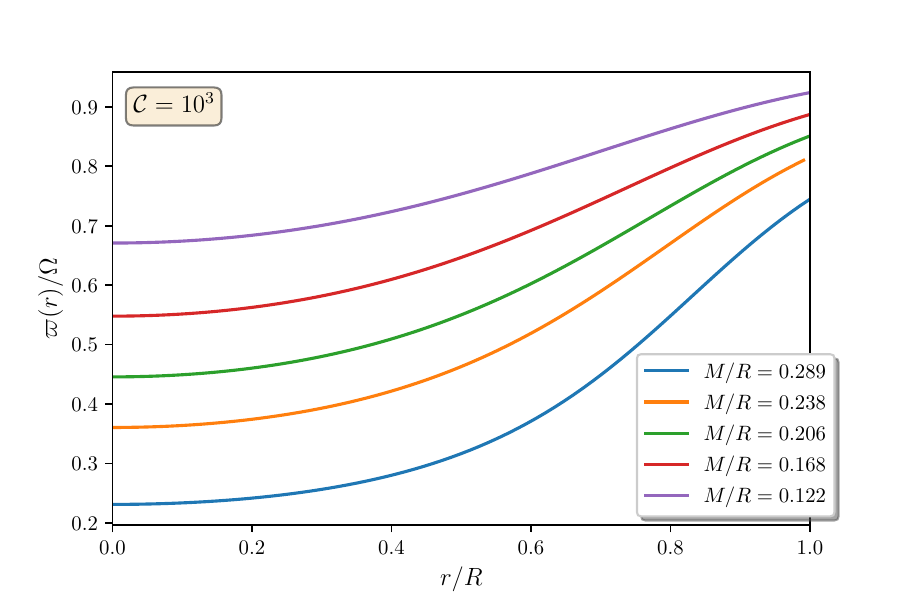}
\includegraphics[width=.495\linewidth]{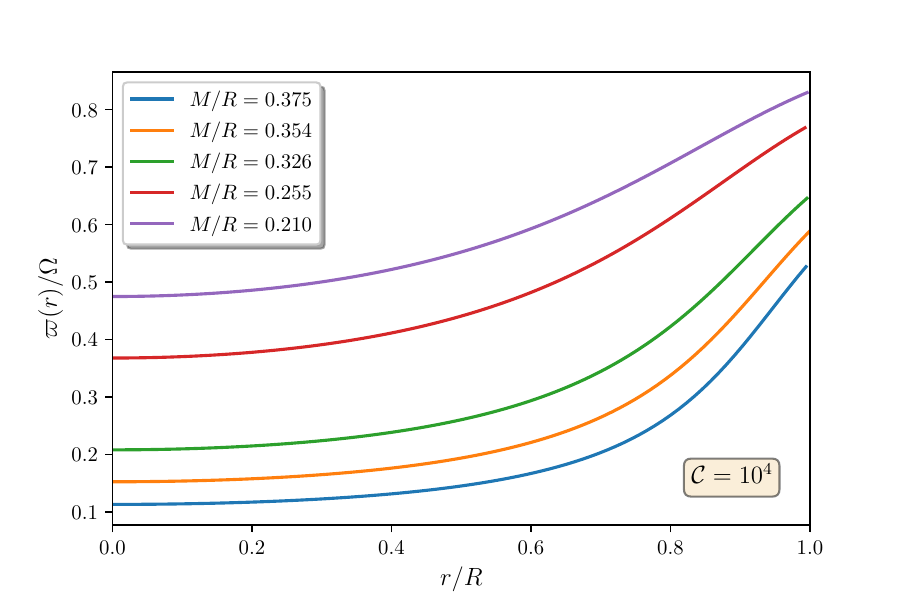}
\caption{Radial profiles of the $\varpi$ function, in the unit $\Omega$, for different compactness, for various values of the anisotropy parameter $\mathcal{C}$. Observe that as the compactness increases, $\varpi$ at the origin decreases. Notice that the highest compactness configurations we consider here are determined by the $M(R)$ stability criteria, so we do not have configurations especially close to the limiting compactness when the central pressure diverges for that particular EOS, nor to the BH limit.}
\label{fig:wnr}
\end{figure*}

Following the conventions from Ref.~\cite{Beltracchi:2024dfb}, we measure $\varpi$ in units of $J/\Rs^3$, where $\Rs\equiv 2GM/c^2$ is the Schwarzschild radius. Thus it is convenient to introduce the quantity $\widetilde{\varpi}\equiv\varpi/(J/\Rs^3)$. In Fig.~\ref{fig:wnr} we display the variation of the function $\varpi$ through the body, in units of $\Omega$, for different configurations. Following Ref.~\cite{Raposo:2018rjn} we consider the anisotropy parameter values: $\con=\{0, 10^2, 10^3, 10^4\}$. We observe that as the compactness increases, $\varpi$ at the origin approaches zero. We point out that the configurations we are considering here are always below the critical compactness for stability, as given by the $M(R)$ method.     

\begin{figure*}
\centering
\includegraphics[width=.495\linewidth]{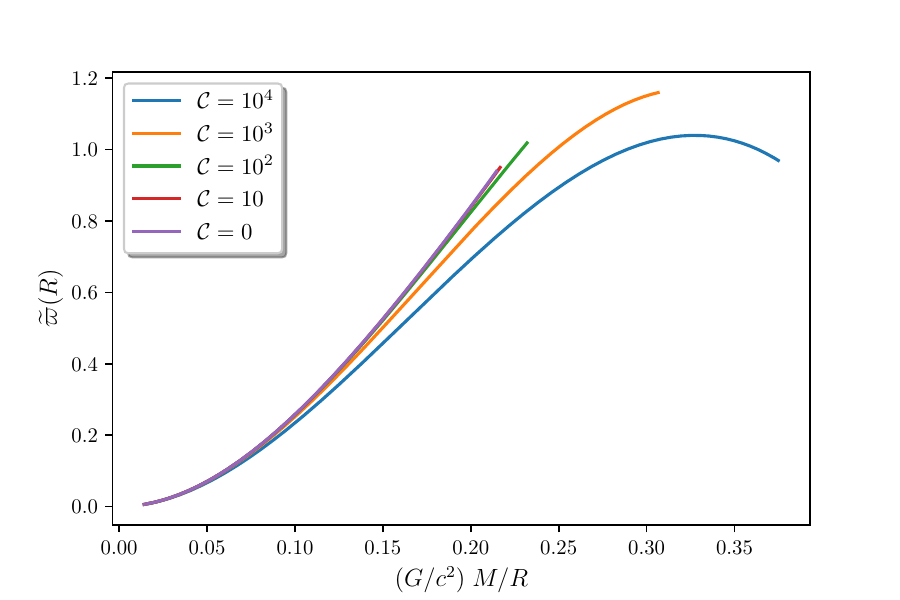}
\includegraphics[width=.495\linewidth]{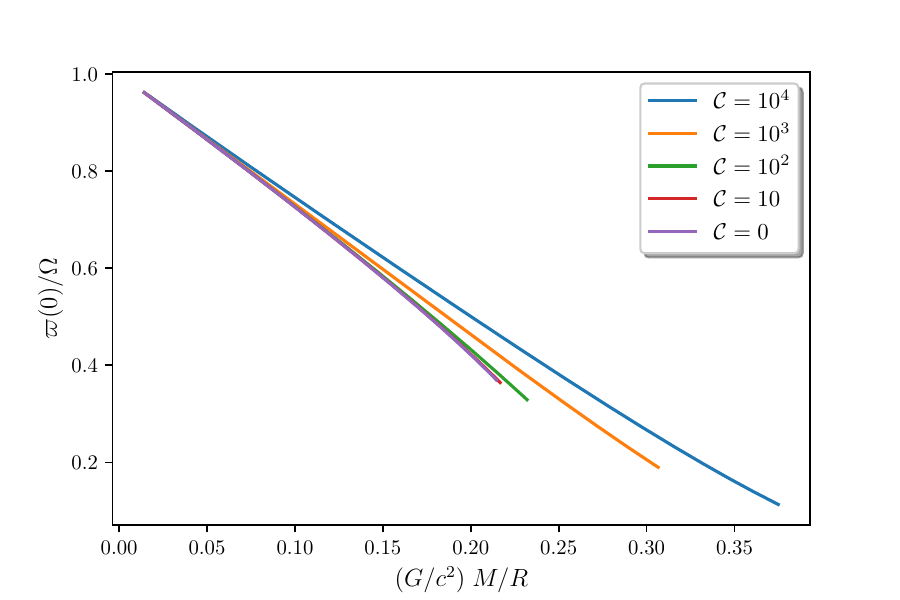}
\caption{{\bf Left panel:} surface value of the $\widetilde{\varpi}$ function, as a function of compactness, for various values of the anisotropy parameter $\con$. {\bf Right panel:} function $\varpi$ evaluated at the origin, as a function of the compactness, for the same values of $\con$ as in the left panel. We measure $\varpi(0)$ in the unit $\Omega$. Observe that at any given anisotropy, $\varpi(0)$ decreases with compactness, and at any given compactness $\varpi(0)$ increases with anisotropy.}
\label{fig:wns}
\end{figure*}

In the left panel of Fig.~\ref{fig:wns}, we show the surface value $\widetilde{\varpi}(R)$, as a function of compactness, for various values of the anisotropy parameter $\con$. For low compactness, we observe that $\widetilde{\varpi}(R)\to 0$, implying $\omega\approx \Omega$, as expected in the Newtonian limit. The right panel of the same figure shows the central value $\varpi(0)$, in units of $\Omega$, as a function of compactness for the anisotropy values as in the left panel. For a given anisotropy, as the compactness increases, $\varpi(0)$ decreases. Notably, in highly anisotropic configurations, $\varpi(0)$ appears to approach zero, suggesting that in such cases $\omega(0)$ approaches the angular velocity of the fluid $\Omega$. This same behavior was also observed in Schwarzschild stars at the Buchdahl limit~\cite{Beltracchi:2023qla} and in Bowers-Liang spheres~\cite{Beltracchi:2024dfb}. However, since our analysis is constrained by stability criteria, we do not explore configurations near the point where pressure diverges, ensuring that $\varpi(0)$ remains strictly above zero in all cases considered.

\begin{figure}
\centering
\includegraphics[width=\columnwidth]{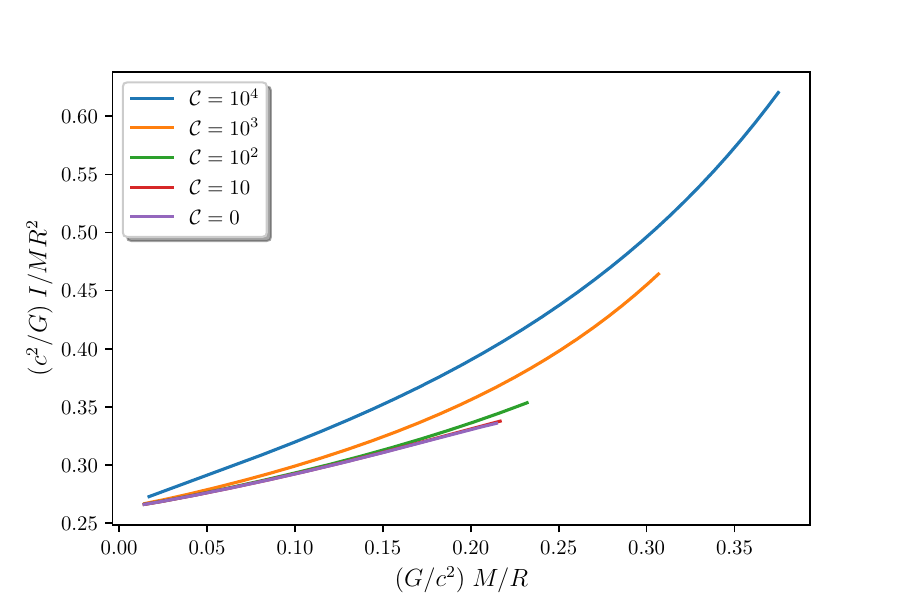}
\caption{Normalized moment of inertia $I/MR^2$ of $\con$-stars, as a function of the compactness, for various values of the anisotropy parameter $\con$.}
\label{fig:inertia}
\end{figure}

In Fig.~\ref{fig:inertia}, we present the normalized moment of inertia $I/MR^2$, as a function of the compactness for varying values of $\con$. In the isotropic case ($\con=0$), our results show excellent agreement with those reported by Ref.~\cite{Miller:1977} for slowly rotating polytropes with $\gamma=2$. In the low compactness limit, we observe that all the curves converge to the Newtonian value $I/MR^2=2(\pi^2-6)/3\pi^2=0.261$.\footnote{This result comes from solving the Newtonian hydrostatic equilibrium equation, for an isotropic polytrope with $n=1$.} For fixed compactness, an increase in the anisotropy parameter $\con$ leads to an increase in the normalized moment of inertia. While one may expect that as anisotropy increases, the maximum allowed compactness approaches the BH limit\textemdash $I/MR^2\to 1$\textemdash the solutions considered here are constrained by the stability criterion. Consequently, the strict BH limit cannot be explored within the scope of these configurations. The moment of inertia of $\con$-stars was also reported by Ref.~\cite{Becerra:2025snm}.

\subsection{Monopole sector}

The monopole sector is governed by Eqs.~\eqref{newm0}, \eqref{newh0}, and \eqref{newp0}, involving the metric perturbation functions $m_0$, and $h_0$, and the EMT eigenvalue perturbation functions $\EE_0$, $\PP_0$, and $\TT_0$. The exterior spacetime depends on the parameter $\delta M$ [see Eq.~\eqref{m0out}], which is determined by the monopole sector. The system's monopole deformation is given by $\xi_0$.

\subsubsection{Behavior of the monopole perturbation functions}

We can draw some observations from the EOSs governing the $\con$-stars, and the EOSs for the perturbations, namely, Eqs.~\eqref{PPvsZZ}, \eqref{EEvsZZ}, and \eqref{TTvsZZ}. The perturbation functions $\PP_x$ and $\TT_x$ typically go to zero at the surface, as both $p_r$ and $\rho_0$ vanish there. On the other hand, $\EE_x$ and $\mathcal{Z}_x$ tend to the same finite value at the boundary. This also leads to drawing some conclusions about the deformation $\xi_0$. If the rest mass density at the surface goes to zero, at least linearly, i.e., 
\begin{equation}
\rho_0(r\approx R^-)=\rho_0^a(R-r)+\cdots, 
\label{rho0surface1}
\end{equation}
then 
\beq
p_r(r\approx R^-) = \kappa\rho_0^a(R-r)^2+\cdots,
\eeq
thus $p'_r(r\approx R)=-2\kappa\rho_0^a(R-r)$. This implies that $\xi_0=-\PP_0/p'_r(r\rightarrow R^-)$ would formally be an ill-defined, and problematic for numeric computation, $0/0$ form. However, using Eq.~\eqref{rho0surface1} for the behavior near the surface, as well as the postulated expansions $m(r\approx R^-)=M+m_a(R-r)+\cdots$, and, $\nu_0(r\approx R^-)=\nu_R+\nu_a(r-R)+\cdots$, the background Einstein equations \eqref{atov}--\eqref{dnudr} and the EOSs \eqref{raposoeos} and \eqref{polytropdef}, with $\kappa=100$, and $n=1$, imply that a solution exists with 
\begin{equation}
\rho_0^a=\frac{M}{200R(R-2M)},\label{rho0asurf}
\end{equation}
therefore $p_r\approx\frac{M^2}{400R^2(R-2M)^2}(r-R)^2 $ and
\begin{equation}
p'_{r}(r\approx R^-)=\frac{M^2}{200R^2(R-2M)^2}(r-R)+\cdots.\label{prpsurf}
\end{equation}
Using Eq.~\eqref{PPvsZZ}, together with Eqs.~\eqref{rho0surface1}, \eqref{rho0asurf}, and \eqref{prpsurf}, we can explicitly divide the lowest-order terms in the series expansions to obtain
\beq
\begin{split}
\xi_x &= -\frac{\PP_x(R)}{p_r'(r\rightarrow R^-)}\\
    &=-\frac{k\gamma \rho_0^{\gamma-1} \mathcal{Z}_x}{p'_{r}(r\rightarrow R^{-})}=\frac{200 R(R-2M)}{M}\mathcal{Z}_x(R),
    \label{xiseries}
\end{split}
\eeq
which is readily computable.
\subsubsection{Results for the monopole perturbations}

\begin{figure}
\centering
\includegraphics[width=\columnwidth]{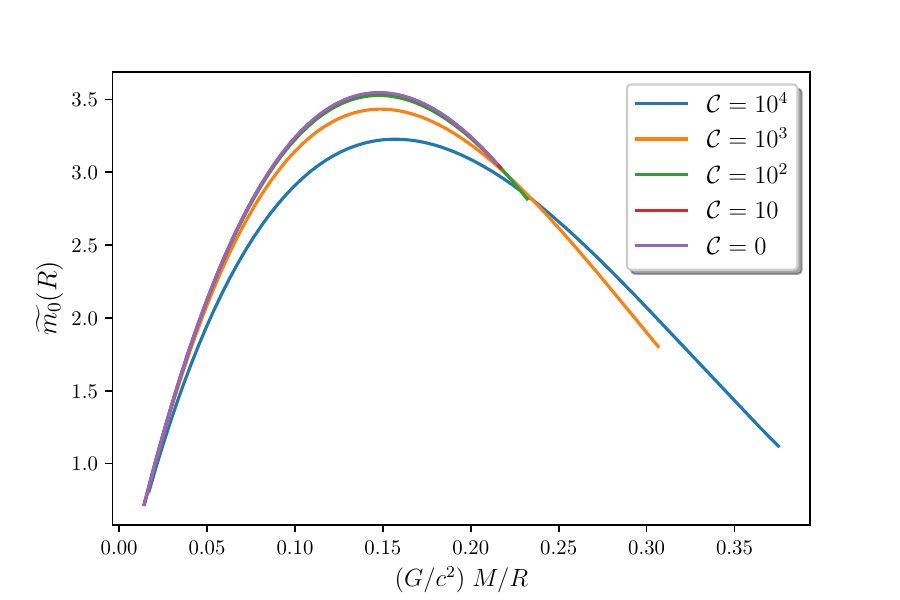}
\caption{Perturbation function $\widetilde{m_0}(R)$, evaluated at the surface, as a function of compactness, for various values of the anisotropy parameter $\con$.}
\label{fig:m0s}
\end{figure}

\begin{figure}
\centering
\includegraphics[width=\columnwidth]{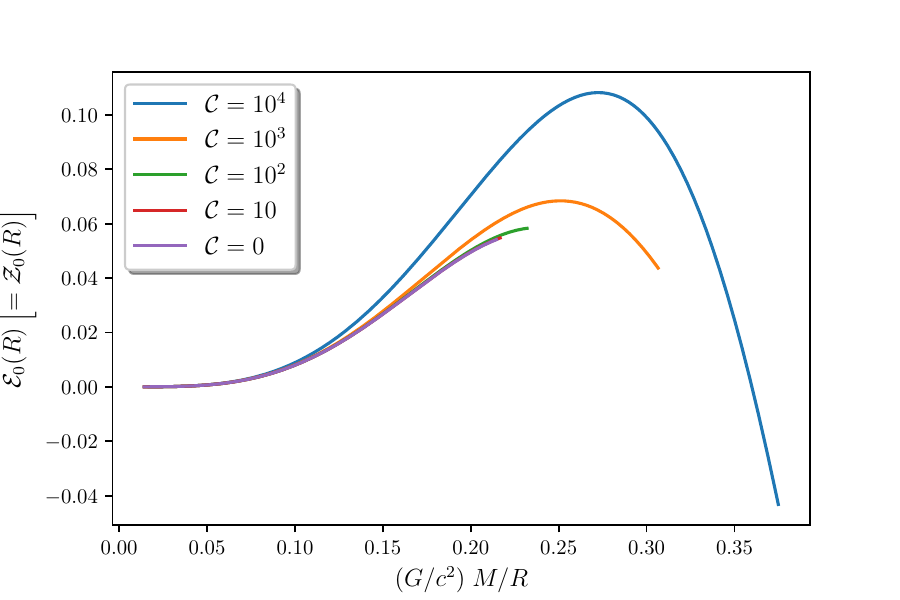}
\caption{Surface value of the perturbation functions $\widetilde{\EE_0}(R)$, and $\widetilde{\ZZ_0}(R)$, against compactness, for various values of the anisotropy parameter $\con$. Note that since the background rest mass density $\rho_0$ vanishes on the surface for the $\con$-star, Eq.~\eqref{EEvsZZ} dictates that the surface values of these perturbation functions are identical.}
\label{fig:p0s}
\end{figure}

Let us now discuss our results for the monopole perturbations of $\con$-stars. To study the spherical deformations, we numerically integrated Eqs.~\eqref{newm0}-\eqref{newp0}, with the initial conditions $h_{0}(0)=m_{0}(0)=\PP_0(0)=0$. Following the conventions of Ref.~\cite{Beltracchi:2024dfb}, it is convenient to define the following quantities
\begin{align}
\widetilde{h_0}&\equiv\frac{h_0}{(J^2/\Rs^4)},& 
\widetilde{m_0}&\equiv\frac{m_0}{(J^2/\Rs^3)},&
\widetilde{\PP_0}&\equiv\frac{\PP_0}{(J^2/\Rs^6)},&
 \nonumber \\
 \widetilde{\EE_0}&\equiv\frac{\EE_0}{(J^2/\Rs^6)},&  \widetilde{\ZZ_0}&\equiv\frac{\ZZ_0}{(J^2/\Rs^6)}.
\end{align}
In Fig.~\ref{fig:m0s}, we display the surface value of the perturbation function $\widetilde{m_0}$, as a function of compactness, for various values of $\con$. Observe that the various curves show a local maximum, then $\widetilde{m_0}$ decreases as the compactness increases.
Figure~\ref{fig:p0s} shows the perturbation functions $\widetilde{\EE_0}(R)$ and $\widetilde{\ZZ_0}(R)$, evaluated at the surface, for varying compactness. Note that, according to Eq.~\eqref{EEvsZZ}, both functions approach the same value at the surface, since the rest mass density vanishes there. We consider the same values of the anisotropy parameter as in Fig.~\ref{fig:m0s}. For the curves $\con=\{0,10,10^2\}$, observe that the perturbation functions increase with compactness. A slightly different behavior is observed in the cases $\con=\{10^3,10^4\}$, where the curves show a local maximum, and then decrease before reaching the critical point for instability. One may suspect that the maximum of the curves with low $\con$ will be found beyond the point of instability; however, we remark that this regime cannot be explored within these solutions.

It is worthwhile to mention that for $n=1$ and $\kappa=100$, the perturbations $\mathcal{P}_x$ and $\mathcal{T}_x$ typically vanish at the surface, as they are proportional to $p_r$ and $\rho_0$ which also go to zero at the surface. In contrast, $\mathcal{E}$ and $\mathcal{Z}$ approach the same finite value.

\begin{figure}
\centering
\includegraphics[width=\columnwidth]{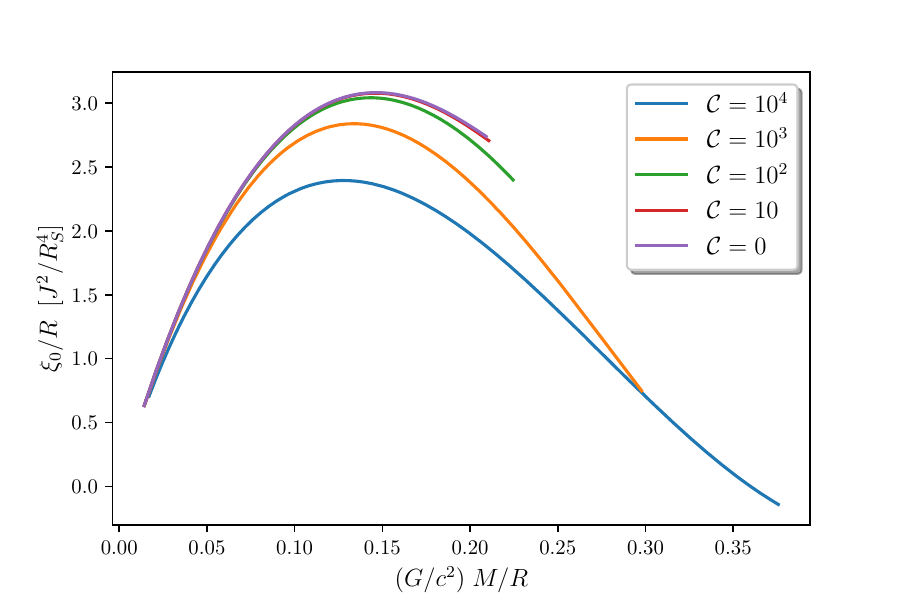}
\caption{Surface values of the spherical deformation function $\xi_0/R$, as a function of compactness, for various values of the anisotropy parameter $\con$. We measure $\xi_0/R$ in units of $J^2/\Rs^4$. }
\label{fig:xi0}
\end{figure}

The $l=0$ deformation of the boundary $\xi_{0}/R$ is shown in Fig.~\ref{fig:xi0}, as a function of compactness, for varying anisotropy parameter $\con$. The function $\xi_{0}/R$ is measured in units of $J^2/\Rs^4$. Observe that the various curves show a local maximum. It is noteworthy that for the highly anisotropic case $\con=10^4$, $\xi_{0}/R$ transitions to negative values beyond $M/R\sim 0.35$. This is remarkably similar to highly anisotropic Bowers-Liang spheres~\cite{Beltracchi:2024dfb} where we found the same behavior and a subsequent approach to zero.

\begin{figure*}
\centering
\includegraphics[width=\columnwidth]{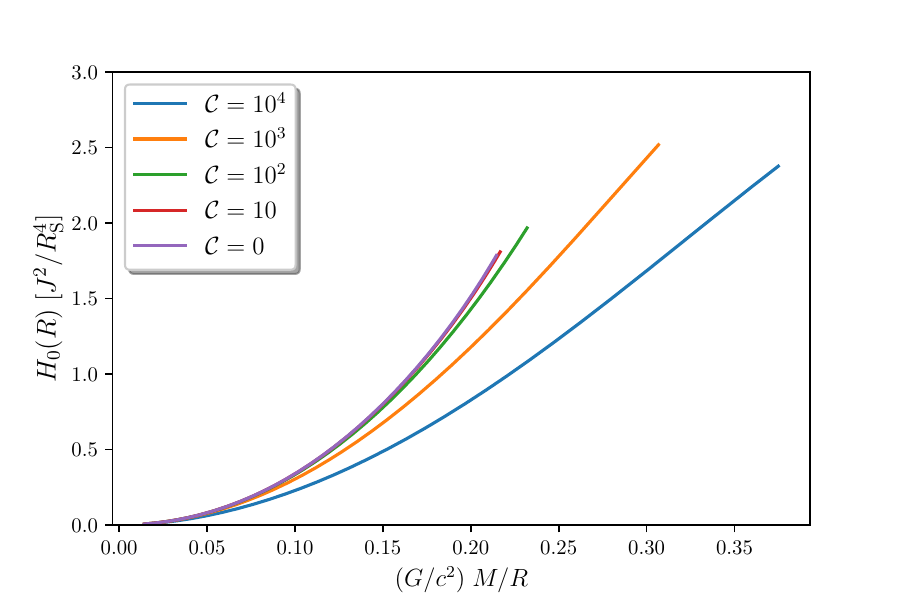}
\includegraphics[width=\columnwidth]{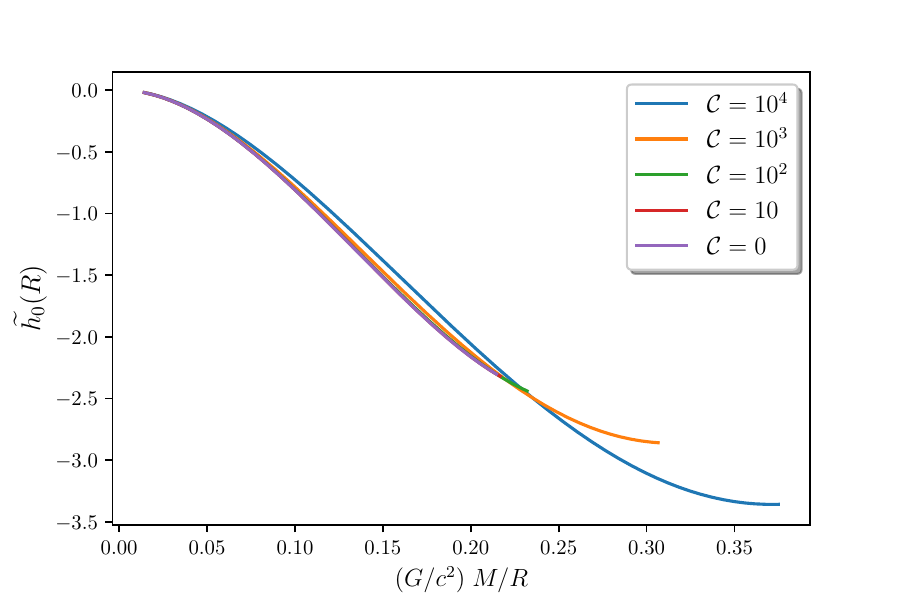}
\caption{{\bf Left:} surface value of the auxiliary function $H_0$ (in units of $J^2/\Rs^4$), as a function of compactness, for various values of anisotropy. For a given value of compactness, as $\con$ increases, $H_0(R)$ decreases. {\bf Right:} surface value of the perturbation function $\widetilde{h_0}(R)$, as a function of compactness, for the same values of the anisotropy parameter $\con$ as in the left panel.}
\label{fig:h0s}
\end{figure*}

As it was discussed in Ref.~\cite{Beltracchi:2024dfb}, the computation of the true perturbation function $h_0$ requires a few steps. First, one finds $H_0$ as a solution to Eqs.~\eqref{newh0} and \eqref{hc_eqn}, assuming a certain integration constant. Then, one determines the corresponding integration constant, such that the interior solution matches the exterior solution given by 
\beq
h_0 = \frac{1}{r-2M}\left(\frac{J^2}{r^3}-\delta M\right),\quad r>R.
\eeq
Since the background $g_{tt}$ is finite and nonzero, there will be no mismatch with the $h_0$ functions. Figure~\ref{fig:h0s} displays the perturbation functions $H_0$ and $h_0$, evaluated at the surface, for various anisotropy values $\con$. Observe that, for a fixed compactness, as the anisotropy increases $H_0(R)$ decreases.

\begin{figure}
\centering
\includegraphics[width=\columnwidth]{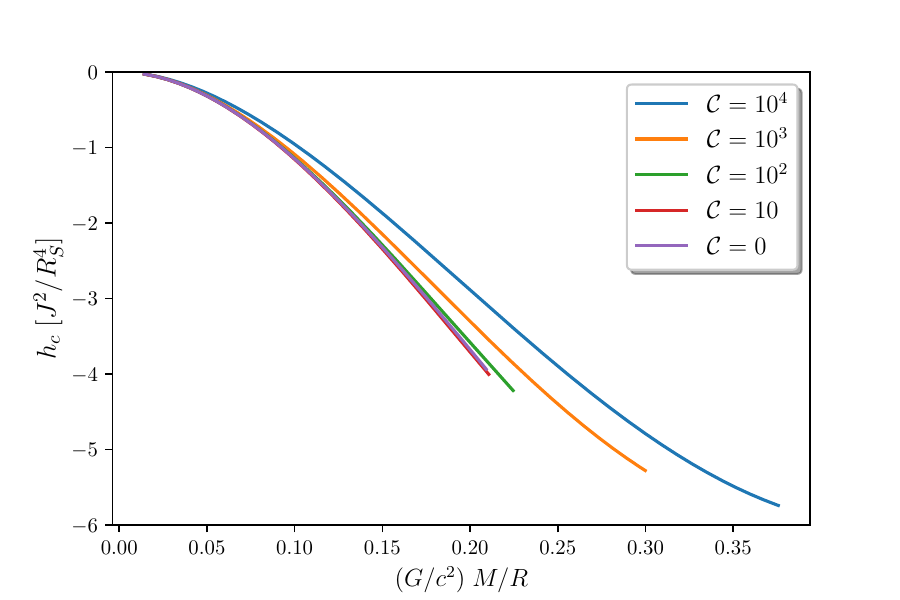}
\caption{Profiles of the constant $h_c$, as a function of compactness, for various values of the anisotropy parameter $\con$. We measure $h_c$ in units of $J^2/\Rs^4$. For highly anisotropic configurations, as the compactness approaches its critical value for stability, $h_c\to -6J^2/\Rs^4$, approximately.}
\label{fig:hc}
\end{figure}

In Fig.~\ref{fig:hc}, we display the profiles of the constant $h_c$, defined by Eq.~\eqref{hc_eqn}, as a function of compactness for varying values of $\con$. The constant $h_c$ is measured in units of $J^2/\Rs^4$. We observe that, for a fixed compactness, $h_c$ increases with anisotropy. Conversely, for a given anisotropy, $h_c$ decreases as compactness increases. For the most compact configurations tested, highly anisotropic systems approach the critical value for instability, with $h_c\to -6J^2/\Rs^4$ approximately. It is noteworthy that this behavior was also observed in sub-Buchdahl Schwarzschild stars in the gravastar limit~\cite{Beltracchi:2023qla}, as well as Bowers-Liang spheres~\cite{Beltracchi:2024dfb}.

\begin{figure}
\includegraphics[width=\columnwidth]{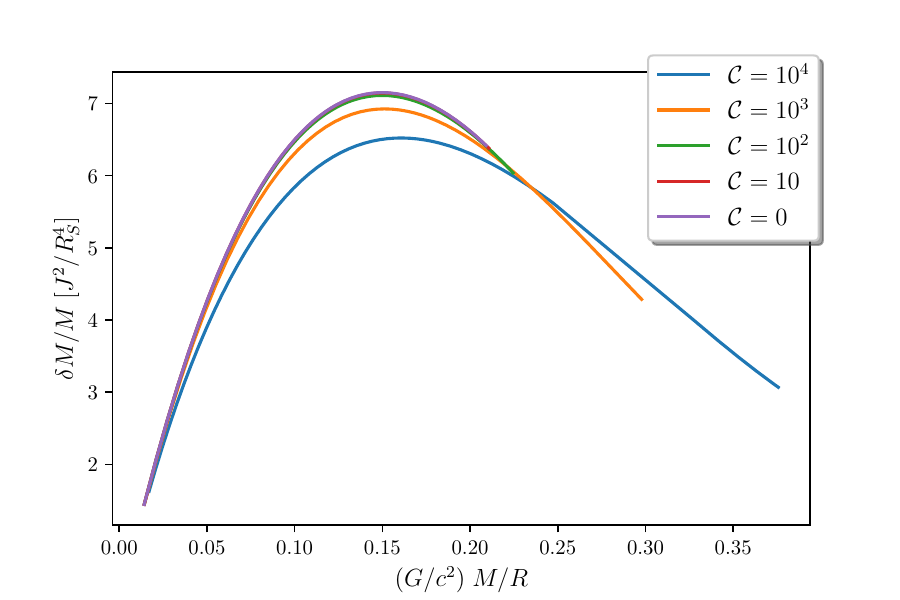}
\caption{Change of mass $\delta M/M$ (in units of $J^2/\Rs^4$), as a function of compactness, for various values of the anisotropy parameter $\con$.}
\label{fig:dM}
\end{figure}

Finally, from Eq.~\eqref{dmMODfin} we determine the change in mass $\delta M$ which we present in Fig.~\ref{fig:dM}, as a function of compactness, for various values of the parameter $\con$. The quantity $\delta M/M$ is measured in units of $J^2/\Rs^4$.

\subsection{Quadrupole sector}

The quadrupole perturbations are determined by Eqs.~\eqref{upsdef}, \eqref{constheta}, \eqref{quadA} and \eqref{quadC}, which features the metric perturbation functions $h_2$, $k_2$, and $m_2$, the EMT eigenvalue perturbations $\EE_2$, $\PP_2$, and $\TT_2$, and the eigenvector perturbation function $\Upsilon$. The surface deformation can be parametrized by $\xi_2$ or $\epsilon$. 

Similarly to the monopole perturbations, the EOSs give that $\PP_2$ and $\TT_2$ vanish at the surface, and $\EE_2$ and $\mathcal{Z}_2$ approach some constant value. The surface deformations $\epsilon$ and $\xi_2$ involve problematic $0/0$ terms, however, Eq.~\eqref{xiseries} allows for $\xi_2$, and hence $\epsilon$ to be computed without any issues.

\subsubsection{Behavior of the quadrupole perturbation functions}

In Ref.~\cite{Beltracchi:2024dfb}, we found that the Einstein equations impose specific constraints on the behavior of perturbations in general well-behaved systems. In particular, the lowest-order terms in $h_2$, $k_2$, $\mathcal{P}_2$, and $\mathcal{T}_2$ behave as $O(r^2)$, while $\Upsilon$ behaves as $O(r)$. Moreover, the leading coefficients of the perturbations $\mathcal{P}_2,\mathcal{T}_2,$ and $\Upsilon$ are related to the quadratic and quartic components of $h_2$ and $k_2$. A similar analysis can be performed for the $\con$-star; however, the isotropic and anisotropic cases must be treated separately due to their different behavior near the origin in the background solution.
\paragraph{The isotropic case.}
For the isotropic case, we find that
\begin{align}
h_2^b+k_2^b=\frac{2\pi}{3} \Bigg\{e^{-2\nu_c} \varpi_c^2
   \left[\left(\frac{p_c}{\kappa}\right)^{1/\gamma}+\frac{\gamma~p_c}{\gamma-1
   }\right]\nonumber\\+ k_2^a \left[\left(\frac{p_c}{\kappa}\right)^{1/\gamma
   }+\frac{(3\gamma-2)p_c}{\gamma-1}\right]\Bigg\},
\end{align}
where $\rho_c=\left(p_c/\kappa\right)^{1/\gamma}+p_c/(\gamma-1)$. The surviving lowest-order terms, for well-behaved systems, become
\begin{multline}
\PP_2^{(P)}=\TT_2^{(P)}=\left(k_2^a-\frac{1}{3} e^{-2\nu_c} \varpi_c^2\right)\\
\times\left[\left(\frac{p_c}{\kappa}\right)^{1/\gamma }+\frac{\gamma~p_c}{\gamma-1}\right] r^2+\cdots,    
\end{multline}
\begin{align}
k_2^{(P)}&=k_2^a r^2+\cdots,\\
h_2^{(P)}&=-k_2^a r^2+\cdots,\\
\Upsilon^{(P)}&=0.
\end{align}
Notice that $\PP_2=\TT_2$ and $\Upsilon=0$, as expected for isotropy.
\paragraph{The anisotropic case.}
In the anisotropic case, using series solutions for the background and quadrupole perturbation functions, we obtain
\begin{align}
h_2^b+k_2^b=\frac{4\pi}{3} \Bigg\{e^{-2\nu_c}\varpi_c^2
   \left[\left(\frac{p_c}{\kappa}\right)^{1/\gamma }+\frac{\gamma~p_c}{\gamma-1
   }\right]\nonumber\\-k_2^a \left[\left(\frac{p_c}{\kappa}\right)^{1/\gamma
   }+\frac{p_c }{\gamma-1 }\right]\Bigg\},
\end{align}
such that the surviving lowest-order terms, for general well-behaved systems, become
\begin{align}
    \PP_2^{(P)}&=0,\\
    \TT_2^{(P)}&=-4\left(k_2^a-\frac{\varpi_c^2}{3e^{2\nu_c}}\right)\left[\left(\frac{p_c}{\kappa}\right)^{1/\gamma }+\frac{\gamma~p_c}{\gamma-1}\right] r^2+\cdots,\\
   k_2^{(P)}&=k_2^a r^2+\cdots,\\
   h_2^{(P)}&=-k_2^a r^2+\cdots,\\
   \Upsilon^{(P)}&=-\left(3k_2^a-\frac{\varpi_c^2}{e^{2\nu_c}}\right)\left[\left(\frac{p_c}{\kappa}\right)^{1/\gamma}+\frac{\gamma~p_c}{\gamma-1}\right] r+\cdots.
\end{align}
Note that for the anisotropic $\con$-star, $\PP_2$, similar to $p_r$ (see Appendix~\ref{appendixA}), does not possess quadratic terms in its series expansions near the origin. Interestingly, the anisotropic $\con$-star has cubic terms in the $h_2$ and $k_2$, obeying
\beq
    k_2(r^3)=-h_2(r^3)=\frac{4(\gamma-1)\left(3k_2^a-e^{-2\nu_0}\varpi_c^2\right)}{9\con\left[p_c+(\gamma-1)(\frac{p_c}{\kappa})^{1/\gamma}\right]}.\label{quadcube}
\eeq
\subsubsection{Results for the quadrupole perturbations}
We numerically integrated the quadrupole perturbation functions, Eqs.~\eqref{upsdef}, \eqref{constheta}, \eqref{quadA}, and \eqref{quadC}, from the origin (or rather some cutoff value $r_0=10^{-6}$) up to the surface $R$. For high anisotropies, we had to increase the cutoff value to $10^{-5}$ to avoid numerical instabilities.

Following the conventions employed in Ref.~\cite{Beltracchi:2024dfb} we define the following quantities:
\begin{subequations}
\beq
\widetilde{h_{2}}\equiv \frac{h_2}{(J^2/\Rs^4)},\quad
\widetilde{k_{2}}\equiv \frac{k_2}{(J^2/\Rs^4)},
\eeq
\beq
\widetilde{m_{2}}\equiv \frac{m_2}{(J^2/\Rs^3)},\quad
\widetilde{\PP_{2}}\equiv \frac{\PP_{2}}{(J^2/\Rs^6)},
\eeq
\beq
\widetilde{\TT_2}\equiv\frac{\TT_2}{(J^2/\Rs^6)},\quad
\widetilde{\ZZ_2}\equiv \frac{\ZZ_2}{(J^2/\Rs^6)}.
\eeq
\end{subequations}

\begin{figure}
\centering
\includegraphics[width=\columnwidth]{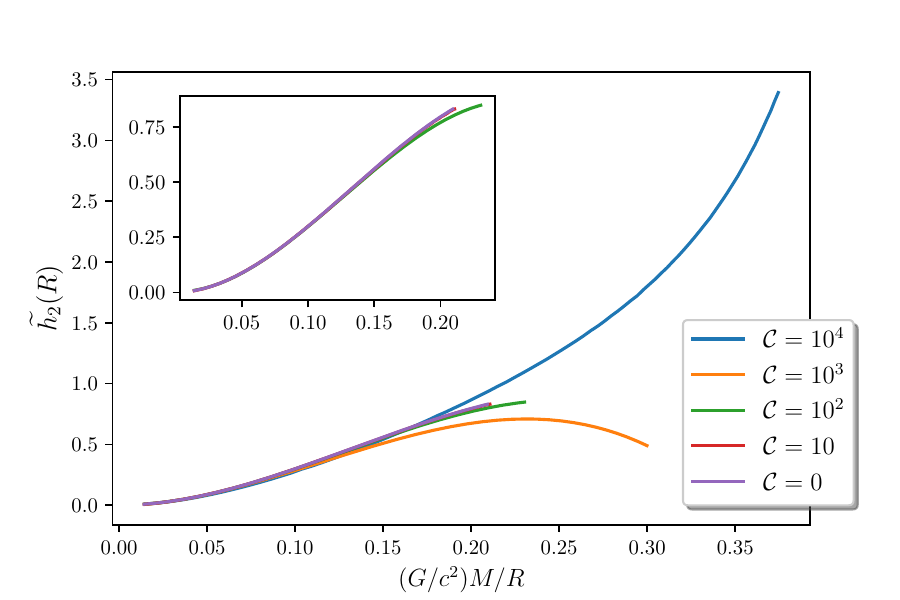}
\caption{Surface value $\widetilde{h_2}(R)$, as a function of compactness, for various values of the anisotropy $\con$.}
\label{fig:h2s}
\end{figure}

In Fig.~\ref{fig:h2s}, we plot the surface value $\widetilde{h_2}(R)$, as a function of compactness, for different values of the anisotropy parameter $\con$. Observe how the curve $\con=10^4$ grows rapidly as the compactness approaches the critical value for instability.

\begin{figure}
\centering
\includegraphics[width=\columnwidth]{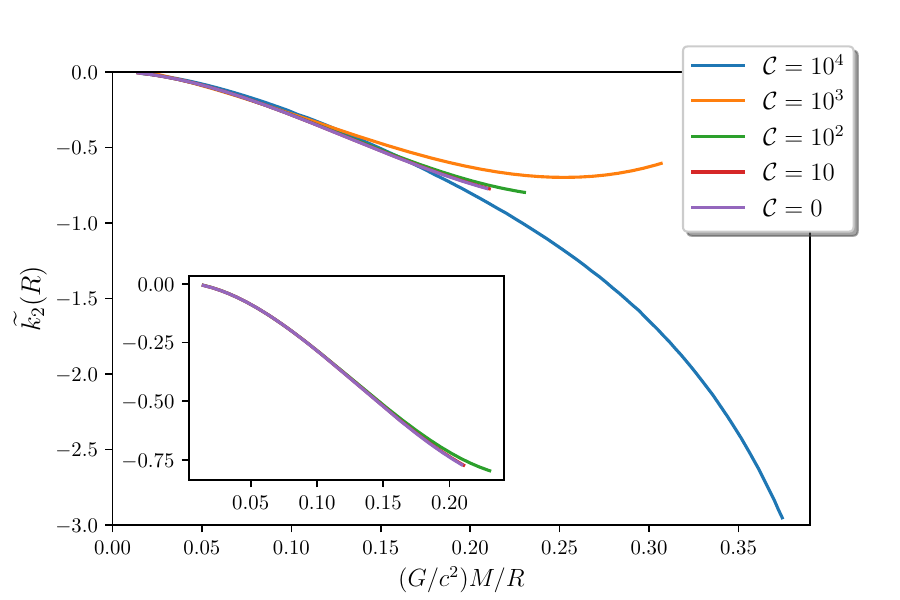}
\caption{Surface value $\widetilde{k_2}(R)$, as a function of compactness, for various values of the anisotropy $\con$.}
\label{fig:k2s}
\end{figure}

In Fig.~\ref{fig:k2s}, we show the surface value $\widetilde{k_2}(R)$, as a function of compactness, for different values of the anisotropy parameter $\con$. Observe that the function $\widetilde{k_2}(R)$ takes negative values. Moreover, the curve $\con=10^4$ decreases rapidly as the compactness approaches the critical value for instability.

\begin{figure}
\centering
\includegraphics[width=\columnwidth]{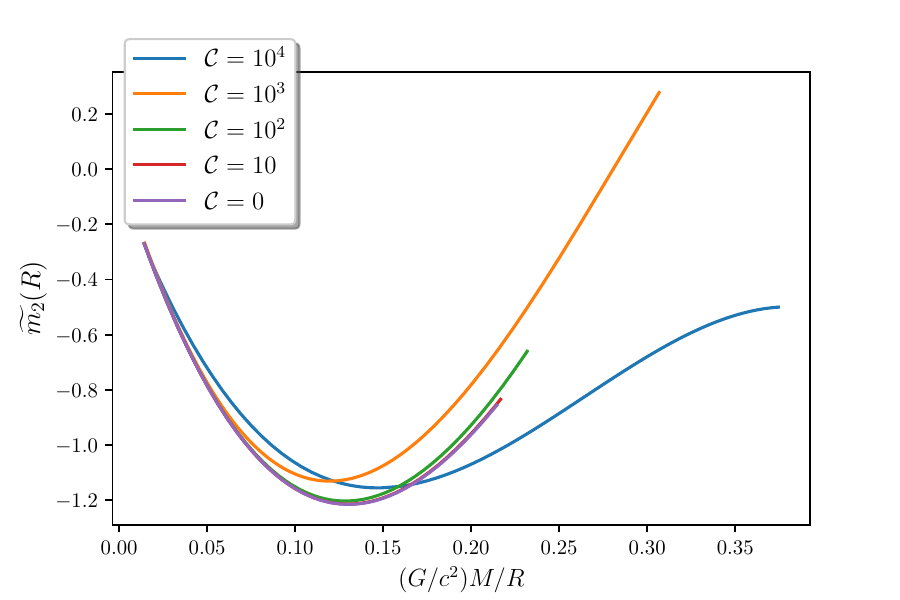}
\caption{Surface value $\widetilde{m_2}(R)$, as a function of compactness, for various values of the anisotropy $\con$.}
\label{fig:m2s}
\end{figure}

Figure~\ref{fig:m2s} displays the value of $\widetilde{m_2}$ at the surface, as a function of compactness, for various values of anisotropy. Observe that the various curves show a local minimum. The curve $10^3$ crosses toward positive values when $M/R\sim 0.28$.

\begin{figure}
\centering
\includegraphics[width=\columnwidth]{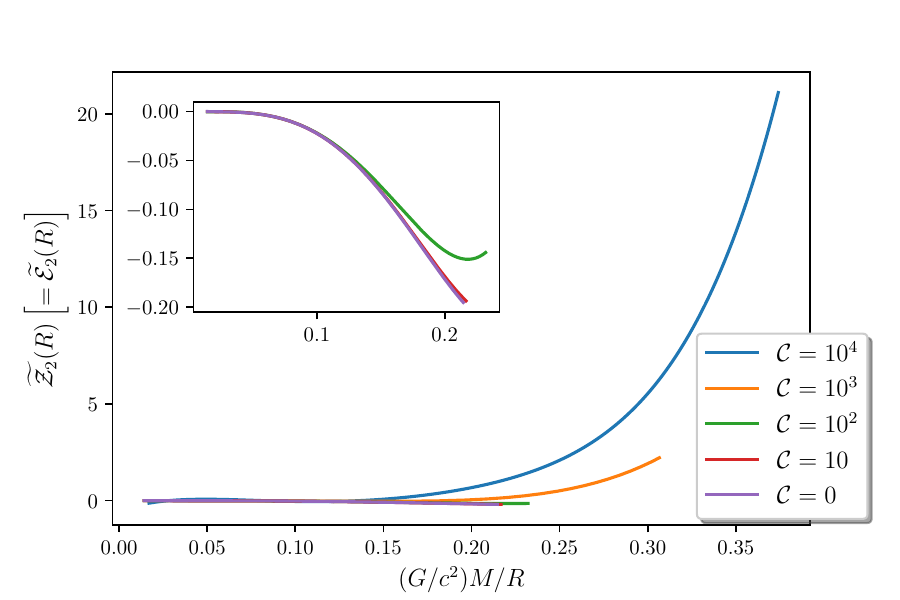}
\caption{Surface value of the perturbation functions $\widetilde{\EE_2}(R)$ and $\widetilde{\ZZ_2}(R)$, as a function of compactness, for various values of the anisotropy $\con$. Recall that the background rest mass density $\rho_0$ vanishes on the surface for the $\con$ star, thus Eq.~\eqref{EEvsZZ} determines that the surface values of these perturbation functions are identical. The inset enhances the curves $\con=\{0,10,10^2\}$.}
\label{fig:z2s}
\end{figure}

Figure~\ref{fig:z2s} shows the surface values of the perturbation functions $\widetilde{\EE_2}(R)$ and $\widetilde{\ZZ_2}(R)$ as a function of compactness for various values of the anisotropy parameter $\con$. The inset highlights the curves corresponding to $\con=\{0, 10, 10^2\}$ to enhance visualization. According to Eq.~\eqref{EEvsZZ}, the surface values of $\widetilde{\EE_2}$ and $\widetilde{\ZZ_2}$ coincide since the rest mass density vanishes at the boundary.

\begin{figure}
\centering
\includegraphics[width=\columnwidth]{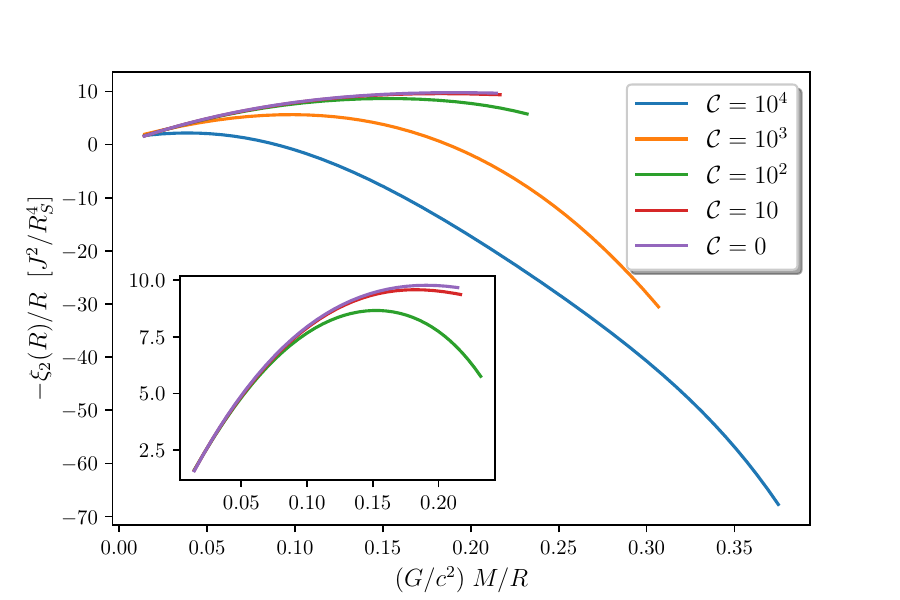}
\caption{The quadrupole perturbation function $-\xi_2(R)/R$, is shown as a function of compactness, for various values of the anisotropy parameter $\con$. The inset provides a detailed view of the curves for $\con=\{0,10,10^2\}$. Observe that the highly anisotropic cases $\con=\{10^3, 10^4\}$, transition to negative values.}
\label{fig:xi2s}
\end{figure}

Figure~\ref{fig:xi2s} displays the quadrupole deformation function $-\xi_2(R)/R$ (in units of $J^2/\Rs^4$), as a function of compactness for various values of the anisotropy parameter $\con$. The inset highlights the curves for $\con=\{0,10,10^2\}$ for better visualization. Notably, while the configurations with $\con=\{0,10,10^2\}$ maintain positive values, the case with $\con=10^3$ transitions to negative values for $M/R\sim 0.18$. Similarly, for the highly anisotropic case $\con=10^4$, the function turns negative when $M/R$ surpasses approximately 0.09. Interestingly, this sign change in the quadrupole deformation function was also observed in highly anisotropic Bowers-Liang spheres~\cite{Beltracchi:2024dfb}.

\begin{figure}
\centering
\includegraphics[width=\columnwidth]{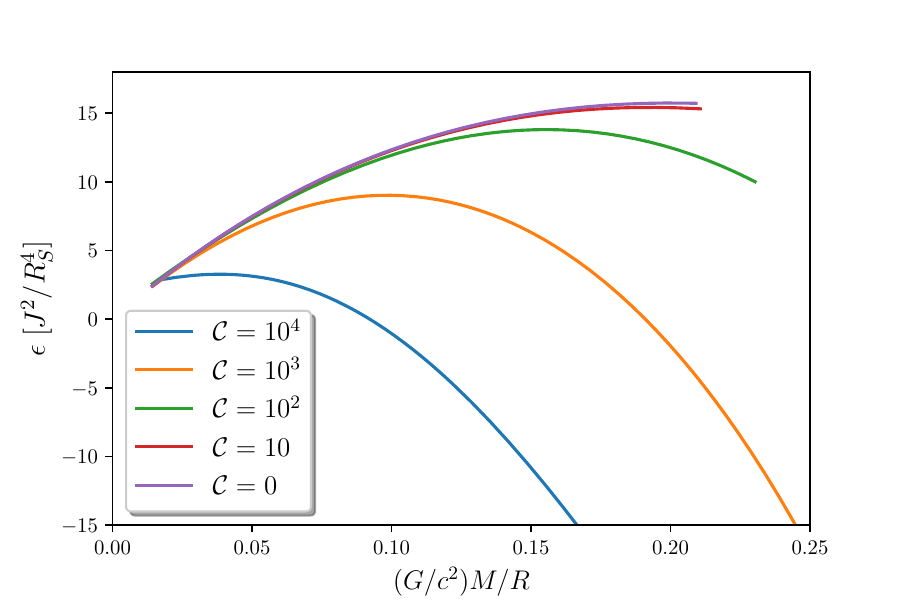}
\caption{Ellipticity of the bounding surface, against compactness, for various values of the anisotropy parameter $\con$. We measure the ellipticity in units of $J^2/\Rs^4$. Observe that the curves $\con=\{10^3, 10^4\}$ transition to negative values of ellipticity (prolate), beyond certain compactness.}
\label{fig:ellip}
\end{figure}

The ellipticity of the boundary, shown in Fig.~\ref{fig:ellip}, follows a similar trend to the $-\xi_2$ curves. With the exception of the isotropic case, all curves exhibit a maximum, and higher anisotropy curves lie below those with lower anisotropy. Notably, for highly anisotropic cases, the ellipticity takes negative values, indicating that these configurations become prolate rather than oblate. Although the authors of Ref.~\cite{Becerra:2024xff}, in their study of slowly rotating piecewise polytropes with a quasilocal anisotropic EOS, did not analyze the ellipticity of these configurations, they found that the mass quadrupole moment changes sign for high anisotropies and compactness. Based on this, they concluded that such configurations are prolate. Thus, these findings, along with our results, suggest that highly anisotropic systems frequently tend to be prolate rather than oblate. In Appendix~\ref{appendix B}, we examine the possible origin of this behavior by analyzing the correspondence of the relativistic energy conservation equation to classical force densities. It is worthwhile to mention that we obtained a slight difference between the various codes, in the surface values of $\epsilon$ and the $\xi_2$ function for $\con=\{10^3, 10^4\}$ at high compactnesses due to the rapidly growing behavior of the function $\ZZ_2$ in the vicinity of the star surface.

\begin{figure}
\centering
\includegraphics[width=\columnwidth]{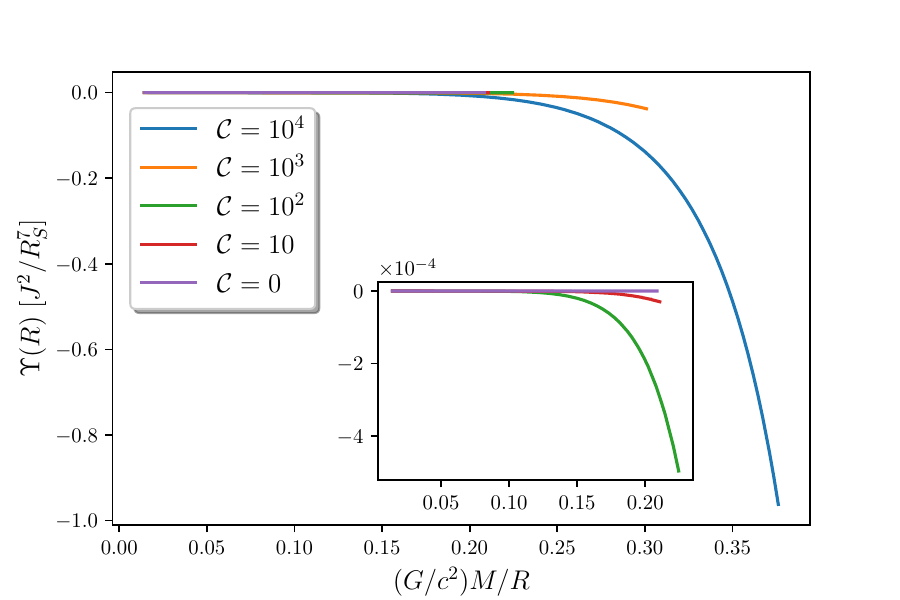}
\caption{Profiles of the auxiliary function $\Upsilon$, evaluated at the surface, as a function of compactness. We plot different values of the anisotropy parameter $\con$. The inset provides an enhanced view of the curves $\con=\{0,10,10^2\}$. Observe that in the isotropic case $\con=0$, $\Upsilon$ is identically zero.}
\label{fig:ups}
\end{figure}

The surface values of the eigenvector perturbation function $\Upsilon$ are shown in Fig.~\ref{fig:ups}. Notice that $\Upsilon$ is identically 0 for the isotropic case and remains relatively small for most of the cases considered. Moreover, it appears to monotonically become increasingly negative with both compactness and $\con$.

\begin{figure}
\centering
\includegraphics[width=\columnwidth]{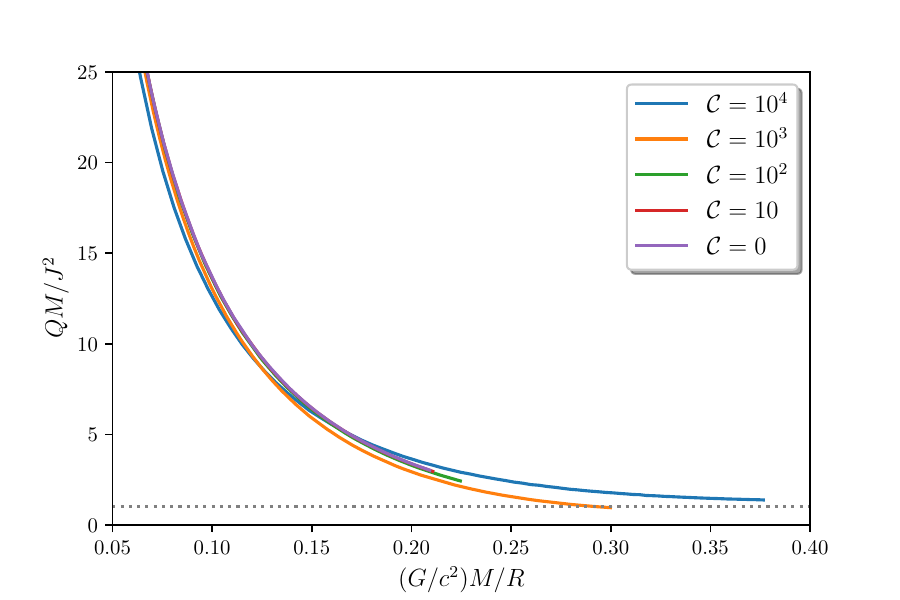}
\caption{Reduced quadrupole moment, or Kerr factor $QM/J^2$, as a function of compactness, for various values of the parameter $\con$. The grey dotted line indicates the Kerr BH value $QM/J^2=1$.}
\label{fig:q}
\end{figure}

While the ellipticity and $\xi_2$ function change sign, for certain configurations, the reduced quadrupole moment, or Kerr factor $QM/J^2$, depicted in Fig.~\ref{fig:q}, remains positive across all tested cases. This indicates that even when the outer boundary exhibits a prolate deformation, in some instances, the overall gravitational effect of the mass distribution still corresponds to an oblate configuration. An intriguing result is that, for $\con=10^3$, the Kerr factor drops slightly below the Kerr BH value (i.e., $QM/J^2<1$) at the critical stability threshold, whereas the $\con=10^4$ curve stays above one (gray dotted line), and appears to approach the Kerr BH value. Additionally, all $Q$ curves decrease monotonically as compactness increases.

\section{Discussion}
\label{sect:disc}
In this work, we investigated the structure and integral properties of slowly rotating anisotropic $\con$-stars, up to the second order in the angular frequency $\Omega$. To this end, we numerically integrated the extended Hartle equations derived in Ref.~\cite{Beltracchi:2024dfb}, varying central mass density and anisotropy parameter. Our study revealed that, in the limit of high anisotropy and maximum compactness, as constrained by stability, certain rotational perturbations of $\con$-stars exhibit a striking similarity to other ultracompact systems. For instance, the constant $h_c$, which appears in the monopole sector [see Eq.~\eqref{hc_eqn}], approaches the value $-6J^2/\Rs^4$, coinciding with the result found for ultracompact Schwarzschild stars in the gravastar limit~\cite{Beltracchi:2023qla}, and highly anisotropic Bowers-Liang spheres~\cite{Beltracchi:2024dfb}. These results suggest that this behavior of the constant $h_c$ may be a generic feature of rotating ultracompact objects. However, further tests on other systems, or possibly a general argument coming from the monopole equations, would be needed to confirm this generality.

Regarding the quadrupole sector, a particularly interesting result is that the ellipticity of the bounding surface changes sign for certain values of anisotropy $\con$, beyond certain compactness. Consequently, in these configurations, slow rotation leads to a prolate rather than an oblate shape. Notably, the ellipticity curves remain smooth and well-behaved across the cases considered, in contrast to the Bowers-Liang model where divergences appeared for specific values of anisotropy and compactness~\cite{Beltracchi:2024dfb}. While the ellipticity transitions to negative values for certain configurations, the reduced quadrupole moment, or Kerr factor, remained positive in all cases studied. Thus, although the surfaces may exhibit negative ellipticity, in certain cases, the overall effect of the quadrupole moment is positive. Finally, we found that for the highly anisotropic case $\con=10^4$, the Kerr factor appears to approach the Kerr BH value, aligning with results from sub-Buchdahl Schwarzschild stars and Bowers-Liang spheres.

The authors of Ref.~\cite{Raposo:2018rjn} concluded that $\con$-stars may provide a prototypical model for ultracompact objects, since they can be as massive and compact as BHs. While we disagree with this conclusion in the context of the configurations studied in this paper, it is plausible that different choices of the parameters $(k,\gamma,\con)$ could yield ultracompact $\con$-stars that more closely approach the BH limit. Although this $\con$-star may provide a more physically viable model for an anisotropic star (in contrast with the Bowers-Liang model), and allows for relatively high compactness, the configurations we examined remained constrained by stability limits well below the BH limit. Therefore, although we observed some indications of certain rotational properties behaving in a similar fashion to ultracompact systems approaching the BH limit, the evidence remains inconclusive. In principle, other ultra-anisotropic $\con$-stars may have limiting compactness very close to the BH value; however, these extremely high values of anisotropy may result in serious numerical instabilities when extended to slow rotation. Thus, identifying other anisotropic systems that support greater compactness, in the stable branch, and examining their behavior under rotation is an area for further investigation.

\section*{Acknowledgments}
We thank the anonymous referees for their comments and suggestions, which helped improve the manuscript. We are also grateful to John C. Miller for valuable discussions.

\appendix

\section{Conditions from the series solutions near the origin}
\label{appendixA}
Suppose $p_r$ can be approximated near $r=0$ as a power series
\beq\label{prexpa}
p_r=p_0+p_1 r+p_2 r^2+p_3 r^3 + \cdots.
\eeq
Substituting Eq.~\eqref{prexpa} into Eq.~\eqref{cstardmpr}, we obtain
\begin{multline}
m = \frac{4}{3}\pi r^3 \left[\left(\frac{p_0}{\kappa}\right)^{1/\gamma }+\frac{p_0}{\gamma-1}\right]\\ 
+\pi r^4\left(\frac{p_1}{\gamma-1}+\frac{\left(\frac{p_0}{\kappa}\right)^{1/\gamma}p_1}{\gamma p_0
   }\right)\\
+\frac{4}{5} \pi r^5 \left[\frac{p_2}{\gamma-1}+\frac{\left(\frac{p_0}{\kappa}\right)^{1/\gamma}\left(p_1^2 - \gamma p_1^2 + 2\gamma p_0 p_2 \right)}{2\gamma^2 p_0^2}\right] + \cdots.
\end{multline}
Using these expressions for $p_r$ and $m$, Eq.~\eqref{cstarTOVpr} results in a series for which we can equate the coefficients term by term. The lowest order term goes like $r^{-1}$ and requires
\beq
\frac{2\left[\con \left(n p_0+\left(\frac{p_0}{\kappa}\right)^{1/\gamma }\right) p_1\right]}{r}=0.
\label{serieslow}
\eeq
To ensure that the pressure does not vanish as we go to the origin, we require $p_0\neq0$. Consequently, this condition leaves us with two possible options: either $p_1=0$ or $\con=0$.
\subsection{The nonzero anisotropy case: $\con\neq0$}
Suppose we have anisotropy, such that $\mathcal{C}\ne0$. The lowest-order term \eqref{serieslow} implies $p_1=0$, and the next lowest $r^0$ term gives
\beq
4\con \left[n p_0+\left(\frac{p_0}{\kappa}\right)^{1/\gamma}\right]p_2=0.
\eeq
Thus, we must have $p_2=0$ if anisotropy is present. However, the next linear in $r$ term gives
\begin{widetext}
\beq
0= \left\{-6\con\left[n p_0+\left(\frac{p_0}{\kappa}\right)^{1/\gamma}\right]p_3 -\left[p_0+np_0+\left(\frac{p_0}{\kappa}\right)^{1/\gamma }\right] \left[4\pi  p_0+\frac{4}{3} \pi \left(\left(\frac{p_0}{\kappa}\right)^{1/\gamma }+\frac{p_0}{\gamma-1}\right)\right]\right\}r,\nonumber
\eeq
\end{widetext}
from which we conclude
\beq
p_3=-\frac{2 \pi \left[p_0+n p_0+\left(\frac{p_0}{\kappa}\right)^{1/\gamma }\right]\left[\left(\frac{p_0}{\kappa}\right)^{1/\gamma }+\left(3+n\right)p_0\right]}{9 \con\left[np_0+\left(\frac{p_0}{\kappa}\right)^{1/\gamma}\right]}.
\eeq
Notice that under the reasonable restrictions $p_0,n,\kappa>0$, we find that $p_3<0$ for $\con>0$, and $p_3>0$ for $\con<0$. However, since we have set $p_1=p_2=0$, a positive $p_3$ implies that the lowest-order nonconstant behavior in pressure leads to an \emph{increase} as we move away from the origin. This contradicts the expected physical behavior, where pressure should \emph{decrease} outward from the origin.

Numerically, for small negative values of $\con$, we find solutions where the pressures remain finite and eventually fall to zero. However, for larger negative magnitudes of $\con$, the solutions exhibit divergences, suggesting that the system becomes unstable, or unphysical beyond a certain threshold. In any case, this outward increase in pressure for negative $\con$ explains why we restrict our analysis to $\con\geq 0$ throughout the paper.\footnote{Note that the authors of Ref.~\cite{Raposo:2018rjn} also restrict their analysis to $\con\geq 0$, though they do not explicitly discuss the reasoning behind this choice.}

\subsection{Isotropic case: $\con=0$}

Suppose we satisfy Eq.~\eqref{serieslow} by considering the isotropic case $\con=0$. The next highest order $r^0$ term still requires
\beq
p_1=0,
\eeq
and the $r^1$ term can be solved to give the constraint
\beq
\begin{split}
p_{2}=-\frac{2\pi}{3}\left[\left(\frac{p_0}{\kappa}\right)^{n/(n+1)} + (n+1)p_0\right]\\
\times\left[\left(\frac{p_0}{\kappa}\right)^{\frac{1}{\gamma}}+\frac{p_0}{\gamma -1}+3p_0\right].    
\end{split}
\eeq
Assuming the same reasonable restrictions of $p_0,n,\kappa>0$, this will always be negative. Consequently, the low-order series terms do not indicate any problematic behavior associated with increasing pressure in the isotropic case.

\section{Analysis of the energy conservation equation and its correspondence with the Newtonian force densities}
\label{appendix B}
The standard TOV equation is 
\beq\label{tovap}
\frac{dp}{dr}=-\left(\rho + p\right)\frac{(m+4 \pi r^3 p)}{r(r-2m)}.
\eeq
It is well known that in the Newtonian limit, Eq.~\eqref{tovap} becomes
\beq
    \frac{dp}{dr}=-\rho \frac{m}{r^2},
\eeq
which relates the gravitational force per volume on the fluid with the pressure gradient, which also has units of force per volume. 

The structure of interpreting the terms of the TOV equation as forces per volume may be extended to the ATOV equation, where the anisotropy term $2(\pt-p_r)/r$, which remains in the Newtonian limit~\cite{Bowers:1974tgi}, also has units of force per volume. Furthermore, suppose spherical symmetry is maintained, but radial collapse or expansion dynamics are permitted. In that case, an additional term involving the change in momentum density shows up in the equation that reduces to the ATOV equation in the static limit, and reduces to an equality between the sum of force densities to the time derivative of momentum densities in the Newtonian limit~\cite{Beltracchi:2018ait}.

The unusual prolate behavior observed in some rotating anisotropic objects can be examined under this line of reasoning. The standard ATOV equation can be most conveniently derived from $\nabla_\mu T^\mu_{~r}=0$ and Eq.~\eqref{dnudr}. For the anisotropic Hartle system, we can write
\begin{align}
   F_P+F_G+F_A+F_R= \nabla_\mu T^\mu_{~r}=0,
\end{align}
where $F_P$ is the pressure gradient term, which explicitly reads
\begin{align}
    F_P=p_r'+\PP_0'+\PP_2'~P_2(\cos\theta),
\end{align}
which can also be written as $\PP'$ and clearly reduces to $p_r'$ when $\Omega=0$. One can see from this term that negative values of $F_P$ correspond to outward forces in this scheme. Conversely, positive values of an $F$ term correspond with inward forces. The quantity $F_G$, which is the generalization of the gravitational force density term, can be written as
\begin{align}
    F_G=&\left[\rho+p_r+\EE_0+\PP_0+(\EE_2+\PP_2)P_2(\cos\theta)\right]\nonumber\\&\times\left[\nu_0'+h_0'+h_2'P_2(\cos\theta)\right].
\end{align}
Notice that $h_0$ and $h_2$ can be seen as perturbations to the potential, and that the first term in the rhs can be written as $(\EE+\PP)$. The cross terms between the perturbation functions are of the fourth order in $\Omega$, which we ignore here. The $F_A$ term is related to anisotropy, and it takes the form

\begin{align}\label{fa}
    F_A=&\frac{2\left[~p_r-\pt+\PP_0-\TT_0+(\PP_2-\TT_2)P_2(\cos\theta)\right]}{r}\nonumber\\&+2\left[e^{\lambda_0}\Upsilon+(p_r-p_t)k_2'\right]P_2(\cos\theta).
\end{align}

The first term of the rhs of Eq.~\eqref{fa} can be written as $2(\PP-\TT)/r$ and contains the background terms that survive when $\Omega=0$. The second term is of the order $\Omega^2$ and has an overall proportionality to the background anisotropy $\pt-p_r$ due to the definition of the eigenvector perturbation function $\Upsilon$ [see Eq.~\eqref{ups_first}]. Therefore, the second term is only present for systems that are \emph{both} perturbed away from the static spherically symmetric case \emph{and} anisotropic. Finally, $F_R$ is related to rotation, and its explicit form is
\beq\label{Fr}
F_R=-e^{-2\nu_0}~r\sin^2\theta~(\rho+\pt)\left[(1-r\nu_0')\varpi^2+r\varpi\varpi'\right].
\eeq
In the Newtonian limit, $\nu_0\approx 0$, $p_\perp<<\rho$, and since the frame dragging function $\omega\approx 0$, then $\varpi\approx\Omega$ and $\varpi'\approx 0$; thus, Eq.~\eqref{Fr} reduces to  
\beq
F_R\rightarrow -r\sin^2\theta~\rho~\Omega^2.
\eeq
Since $\Omega$ is the fluid angular velocity, this is clearly a centrifugal force per unit volume. 
 
We now have various forces to examine regarding the possible origin of the prolate behavior. First, the prolate behavior is seemingly related to the $\theta$ dependent perturbations; thus, anything from the background solution or the monopole perturbation is not suitable.

Prolate deformation has not been observed in isotropic systems, to our knowledge, which makes it plausible that it originates from perturbations to $F_A$. This would either have to originate from the quadrupole perturbation in the first term $2(\PP_2-\TT_2)/r$, or the second term, which is entirely in the quadrupole sector. However, it is possible that the $\PP_2'$ term from $F_P$, or the quadrupole perturbations to the $F_G$, can lead to $\theta$ dependent forces that are inward at the equator.

On the other hand, while the rotation term $F_R$ can be $\theta$ dependent, it tends to be outward (negative) because of the signs of the constituent elements. The terms $e^{-2\nu_0}$,  $(\rho+\pt)$, $\varpi$, and $\varpi'$ are separately nonnegative for the well-behaved systems we have examined. The remaining element $(1-r\nu_0')=(r-3m-4\pi r^3 p_r)/(r-2m)$ is typically positive, although it can be negative for systems with high compactness, or large pressure. One such example is near the surface of a star with $M/R>1/3$. Even in these cases, there is no guarantee that $F_R$, as a whole, would change sign because $(1-r\nu_0')\varpi^2$ is still added to the positive term $r\varpi\varpi'$. Furthermore, $(1-r\nu_0')$ can become negative even for isotropic systems, and is not known to cause prolate behavior in such cases. For instance, no prolate behavior was found in the Schwarzschild star in the range $4/9>M/R>1/3$, which is the regime between the sign change in $(1-r\nu_0')$ on the surface and the Buchdahl limit \cite{Chandra:1974,Beltracchi:2023qla}. Finally, prolate behavior, in the form of a negative ellipticity of the boundary, is observed for highly anisotropic $\con$-stars with compactness $M/R<1/3$ where the net effect of $F_R$ must be outward (see Fig.~\ref{fig:ellip}).

\bibliography{ref}
\bibliographystyle{apsrev4-1}

\end{document}